\documentclass[sigconf]{acmart}
\AtBeginDocument{%
  }

\usepackage{amsfonts}
\usepackage{textcomp}
\usepackage{xcolor}
\usepackage{graphicx,epsfig,amsmath,url,latexsym}
\usepackage{ifthen}
\usepackage{verbatim}
\usepackage{paralist}
\usepackage{xspace}
\usepackage{caption}
\usepackage{subcaption}
\usepackage{cases}
\usepackage{bm}
\usepackage{balance}
\usepackage{algorithm} 
\usepackage{algpseudocode}
\usepackage{multirow}
\usepackage{url}
\usepackage{enumitem}
\usepackage{float}
\usepackage{hyperref}
\usepackage{xspace}
\usepackage{caption}

\newcommand{\sysname}{VReaves\xspace}

\setcopyright{none}
\renewcommand\footnotetextcopyrightpermission[1]{}  
\begin{document}

\title{\sysname: Eavesdropping on Virtual Reality App Identity and Activity via Electromagnetic Side Channels}

\author{Wei Sun$^{1}$, Minghong Fang$^{2}$, Mengyuan Li$^{3}$}
\email{redsunwit@gmail.com,  minghong.fang@louisville.edu, mengyuanli@usc.edu}
\affiliation{
  \institution{$^{1}$Wichita State University, $^{2}$University of Louisville, $^{3}$University of Southern California}
  \country{}
}

\begin{abstract}
Virtual reality (VR) has recently proliferated significantly, consisting of headsets or head-mounted displays (HMDs) and hand controllers for an embodied and immersive experience. The VR device is usually embedded with different kinds of IoT sensors, such as cameras, microphones, communication sensors, etc. However, VR security has not been scrutinized from a physical hardware point of view, especially electromagnetic emanations (EM) that are automatically and unintentionally emitted from the VR headset. This paper presents \sysname,  a system that can eavesdrop on the electromagnetic emanation side channel of a VR headset for VR app identification and activity recognition. To do so, we first characterize the electromagnetic emanations from the embedded IoT sensors (e.g., cameras and microphones) in the VR headset through a signal processing pipeline and further propose machine learning models to identify the VR app and recognize the VR app activities. Our experimental evaluation with commercial off-the-shelf VR devices demonstrates the efficiency of VR app identification and activity recognition via electromagnetic emanation side channel.
\end{abstract}


\maketitle
\pagestyle{empty}

\section{Introduction}
Virtual reality (VR) systems have become omnipresent in our daily lives, usually consisting of a VR headset or head-mounted display (HMD) and two hand controllers. We can use these VR systems for an embodied and immersive user experience, such as playing video games, watching videos, online training, collaboration, etc. Among all of these, virtual reality gaming~\cite{pallavicini2019gaming,theodoropoulos2022vr} has shown growing interest due to the proliferation of mobile computing and human-computer interaction. As such, there are different kinds of commercial off-the-shelf VR systems being developed on the market, such as Meta Quest~\cite{quest}, Sony PlayStation VR2~\cite{sony}, HTC Vive~\cite{htv}, Apple Vision Pro~\cite{apple}, etc. As a mobile device, the VR headset can run different VR apps, which can reveal the VR user's private information (e.g., personality~\cite{kober2013personality, poushneh2018augmented} and behavior biometrics~\cite{pfeuffer2019behavioural}). These VR platforms are embedded with different types of IoT sensors~\cite{bamodu2013virtual}, such as cameras and microphones for an embodied and immersive user experience, which are vulnerable to different side channel-based privacy attacks~\cite{garrido2023sok,slocum2024doesn,zhang2023s,slocum2023going}.


However, the existing side channel-based attacks on VR systems mainly focus on acoustics~\cite{luo2024eavesdropping}, VR user behaviors (e.g., head movements~\cite{nguyen2024penetration, zhang2023s} or hand gestures~\cite{gopal2023hidden}), RF side channel~\cite{al2021vr}, motion sensors~\cite{nair2023unique}, camera-captured videos/images (e.g., user avatar)~\cite{wang2024gazeploit}, network traffic~\cite{su2024remote}, etc., which leave the electromagnetic side-channel unexploited yet for VR platforms. Moreover, all these attacks either require malware or a proactive surveillance implant in the VR user's environment to obtain private user-related (e.g., motion-related) information. Therefore, it is essential to exploit the automatic and unintentional electromagnetic side channel to scrutinize the VR platform.

\begin{figure}
\centering  \includegraphics[width=0.86\linewidth]{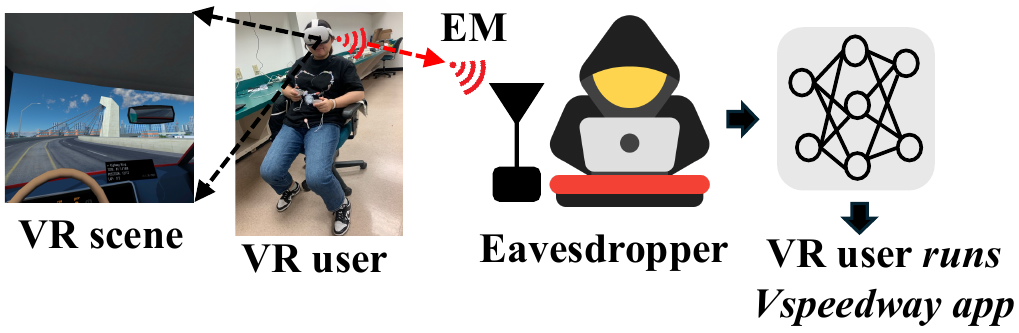}
    \caption{VR headset leaks emanations during operation, which can be sniffed by the eavesdropper to infer the VR app's identities (e.g., Vspeedway app and Aim app) and activities (e.g., app configuration and running).}
    \label{fig:model}
\end{figure}

To this end, this paper presents \sysname, a system that can exploit the electromagnetic emanations (EM) automatically and unintentionally generated by the IoT devices (e.g., camera, microphone) embedded in the VR headset to eavesdrop on the VR headset or head mount display (HMD) as shown in Fig.~\ref{fig:model}, which can reveal important private information (e.g., personalities) related to the VR users through VR app identification and activity recognition. Therefore, this EM side channel can pose a great privacy threat to the VR users. For example, people can make a profit from understanding VR app identities and activities by leveraging this information to make targeted app recommendations and even reveal the user's personality and daily living habits~\cite{poushneh2018augmented, pfeuffer2019behavioural}.

To have functional and practical EM-based eavesdropping on the VR device, the adversary can use a wireless receiver (e.g., software-defined radio) to eavesdrop on the emanations emitted from the VR user's headset for VR app information analysis. However, there are three great challenges. First, even though the emanation as side channel information is widely exploited for hidden camera or microphone detection~\cite{sun2025revealing,zhou2023dehirec,zhang2024eye}, camera image reconstruction~\cite{long2024eye}, and screen information reading~\cite{liu2020screen}, the prior works do not reveal the relationship between the emanations from VR headset and the VR app identities and activities. Different from the prior work of characterizing the single emanation source (e.g., camera or microphone), the VR headset consists of different kinds of emanation sources. As a result, the emanations from these sources are interleaved with each other, which are difficult to extract and characterize using the techniques proposed in the prior works. Therefore, we not only need to accurately extract the frequency-domain emanations but also properly characterize the emanations for the VR app identification and activity recognition.  To do so, we propose to use fine-tuned machine learning models to characterize the relationships between the interleaved emanations and VR app information.

Second, the emanations are unintentionally and automatically emitted from the VR headset, which can be weak in strength and easily interfere with the ambient wireless signals. This is because the emanations are amplitude-modulated clock signals in the time domain that are spreading across a wide frequency band. Therefore, it is important to boost the emanation strength and suppress the ambient wireless interference. To address this challenge,  we propose a signal-processing pipeline, including noise floor smoothing, interference suppression, and emanation strength enhancement to characterize the emanation spectrum accurately. Then, we design a multi-frequency machine learning model to characterize the frequency-domain emanations for VR app identification.

At last, it is difficult to infer the VR app identities and activities based on the same frequency-domain emanations simultaneously. This is because the frequency-domain emanations exhibit the same pattern for a specific VR app, which cannot be used to discriminate the fine-grained app activities. Therefore, we further explore the over-time frequency-domain emanations to characterize the VR app's activities. Specifically, we propose to use short-time frequency transform (STFT) to derive the spectrogram of the emanations and further regard it as an image for the VR app activity recognition using a multi-spectrogram machine learning model.

To demonstrate the efficiency of the emanation-based VR app identification and activity recognition, we built a prototype with the software-defined radio (i.e., USRP N210) instrumented with the LP1401 directional antenna for emanation eavesdropping. Our extensive system performance evaluation with commercial off-the-shelf VR platforms (e.g., Meta Quest 3 and HTC VIVE XR Elite) achieves an accuracy of $99\%$ in the VR app identification and an accuracy of $99\%$ in the VR app activity recognition under various settings including distance, orientation, hardware and software configurations, etc. We summarize the main contribution of our system design in the following. 
\begin{itemize}[leftmargin=*]
    \item To the best of our knowledge, this is the first system that exploits the emanations from the VR headset for VR app identification and activity recognition. 
    \item We propose a signal processing pipeline to smooth the noise floor across a wide frequency band, suppress ambient wireless interference, and boost the emanation strength for accurate emanation extraction and characterization. 
    \item To reveal the relationship between the emanations and VR app identities and activities, we propose to characterize the frequency-domain emanations and over-time frequency-domain emanations through averaging FFT and STFT, using the fine-tuned pre-trained machine learning models.  
    \item Our experimental evaluations with system-level tests, case study, and microbenchmarks demonstrate the efficiency of the VR app identification and activity recognition based on the characterized emanation leakage. 
\end{itemize}
This paper marks an important step in establishing the relationship between the emanations and computational activities in the VR device, which can propel the field forward on VR device scrutiny. Since emanations are modulated by the computational activities, we can leverage these emanations to infer the videos in VR and further reconstruct the VR scenes displayed in the headset. 

\begin{figure}
\centering
    \includegraphics[width=0.5\linewidth]{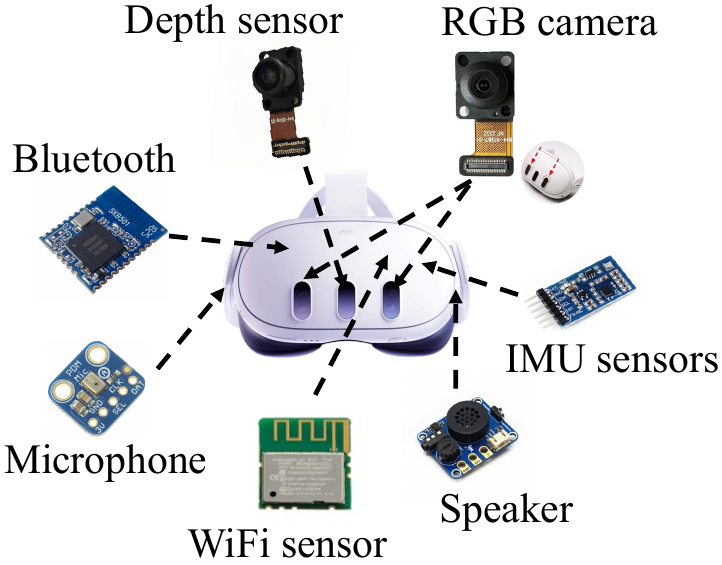}
    \caption{The sensors embedded in the Meta Quest 3 (or HTC Vive XR Elite) include cameras, speakers, microphones, WiFi, and Bluetooth sensors that can be the potential emanation sources.}
    \label{fig:meta:quest:diagram}
\end{figure}


\section{Background}
\label{sec:background}

\subsection{VR Devices and Their Emanations}

\begin{figure}
\centering
    \includegraphics[width=0.85\linewidth]{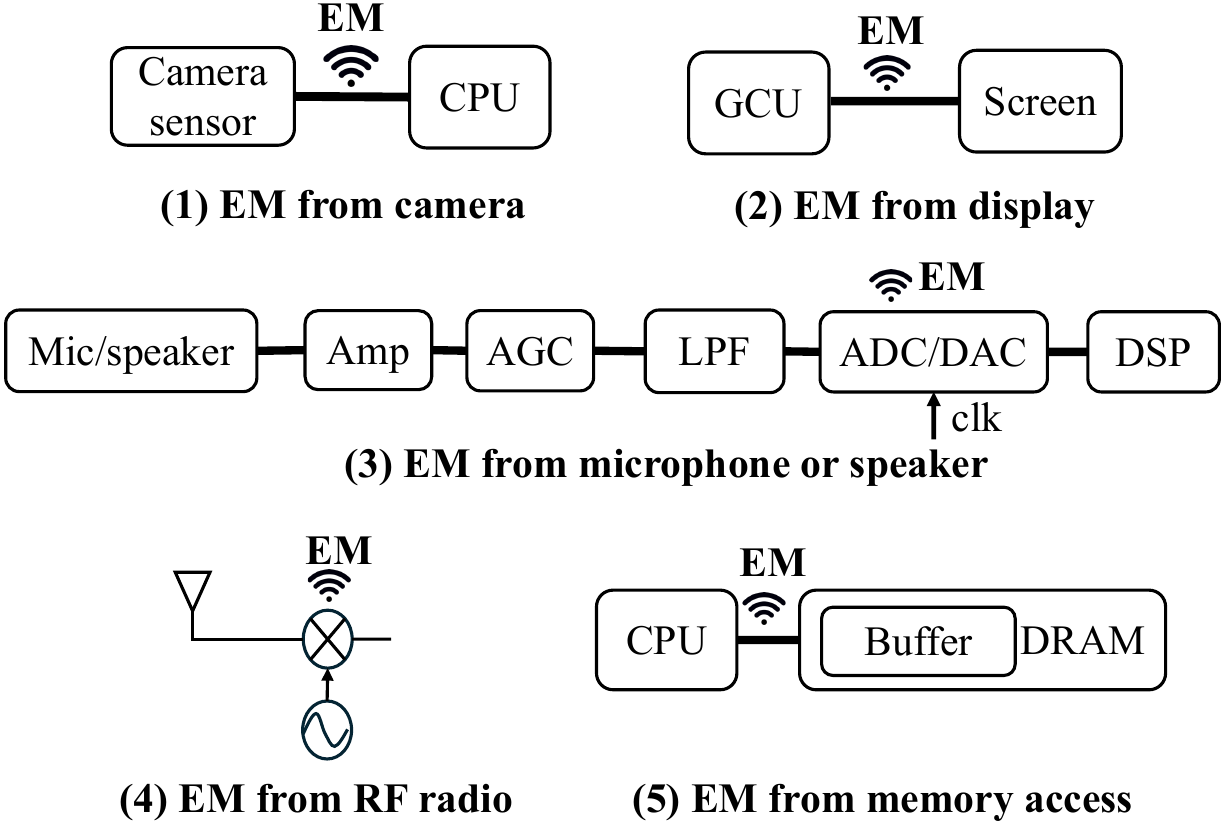}
    \caption{Emanation source in VR headset includes (1) camera sensors, (2) screens or monitors, (3) microphones or speakers, (4) wireless communication radios (e.g., WiFi and Bluetooth sensors), and (5) dynamic random-access memory.}
    \label{fig:prime:em}
\end{figure}

The virtual reality platforms (e.g., Meta Quest, HTC Vive XR Elite, etc.) usually consist of a headset (or head mount display) and two hand controllers.  For example, as shown in Fig.~\ref{fig:meta:quest:diagram},  the Meta Quest 3 headset is embedded with different types of sensors for perception, display, and detection of physical scenes. These embedded sensors can automatically and unintentionally emit electromagnetic emanations, which can be used as a side channel to steal human private information. We primarily summarize five types of emanation sources from the VR headset in the following.
\begin{itemize}[leftmargin=*]
    \item \textbf{Camera's emanations.} The camera sensors are connected to the CPU, GPU, or image processing unit for raw pixel data transmission through High-speed Serial Pixel Interface, Digital Video Port, Low-voltage Differential Signaling, or MIPI Camera Serial Interface 2, which can leak the emanations~\cite{long2024eye,sun2025revealing} that can indicate the computation activities of the camera sensor as shown in Fig.~\ref{fig:prime:em}(1).
    \item \textbf{Display's emanations.} The display in the headset could leak emanations~\cite{long2024eye, liu2020screen}, consisting of the graphical computing unit (GCU) and screen (e.g., OLED or LED displays). The emanations are emitted from the data interface connecting the GCU and screen as shown in Fig.~\ref{fig:prime:em}(2).
    \item \textbf{Microphone and speaker's emanations.} The microphones and speakers can be the emanation sources~\cite{zhou2023dehirec, chen2024eavesdropping}, which usually include the ADC or DAC controlled by the clock signals for synchronization that can leak emanations as shown in Fig.~\ref{fig:prime:em}(3).
    \item \textbf{RF radio's emanations.} The wireless communication radios (e.g., WiFi and Bluetooth radios) can emit emanations~\cite{chaman2018ghostbuster,camurati2018screaming} through the oscillators or mixers as shown in Fig.~\ref{fig:prime:em}(4). 
    \item \textbf{Memory's emanations.} The dynamic random-access memory (DRAM) can introduce the emanations~\cite{shen2021lora, shen2021earfisher, yu2024freeem} in modern electronic devices due to memory access operation of the CPU as shown in Fig.~\ref{fig:prime:em}(5).
\end{itemize}
The emanations are emitted through the circuits and data/signal pipelines on these sensors that can be unintentionally regarded as radio-frequency antennas. Next, we illustrate the primer and properties of the emanations.

\subsection{Primer on Electromagnetic Emanations}

\begin{figure}
\centering
    \includegraphics[width=0.85\linewidth]{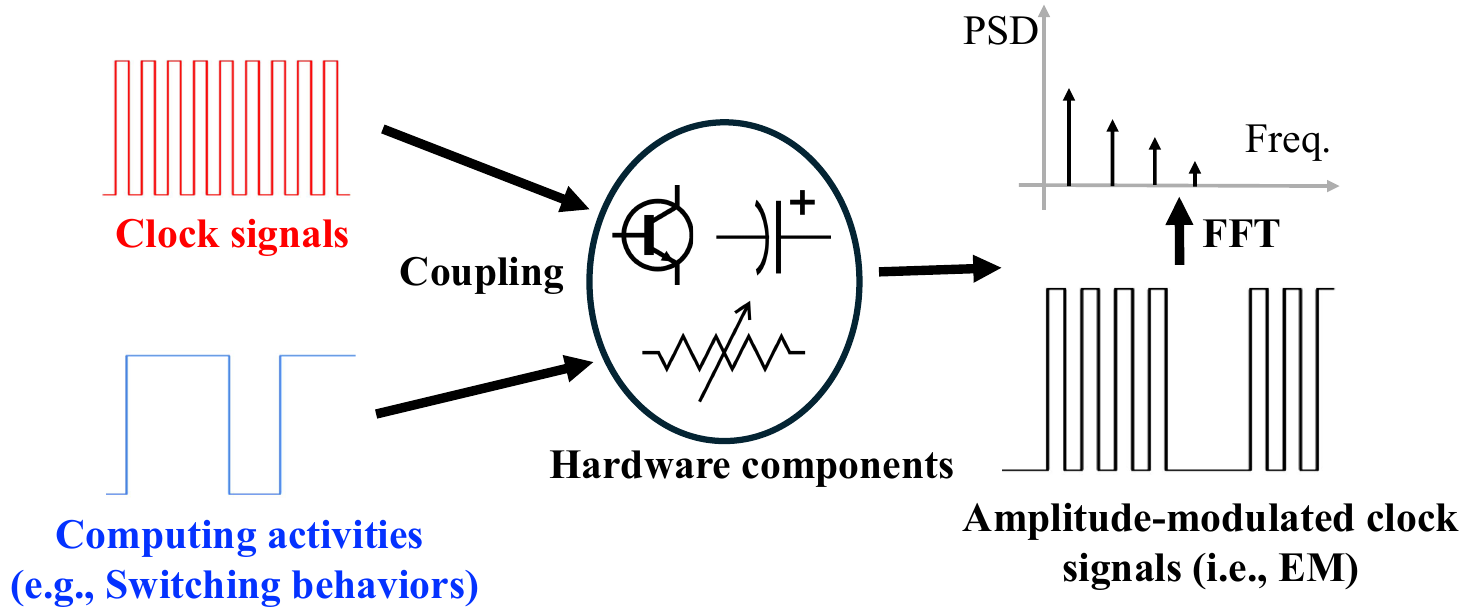}
    \caption{Electromagnetic emanations are amplitude-modulated clock signals, introduced by the clock signals coupling with the computational activities.}
    \label{fig:em:mod}
\end{figure}

\noindent \textbf{Physical principle of EM.} As shown in Fig.~\ref{fig:em:mod}, the electromagnetic emanations (or emanations) are amplitude-modulated clock signals, as the clock signals are coupled with the computational activities (e.g., switching behaviors) through hardware components such as the capacitor, resistor, or diodes. The clock signals can be expressed as follows:
\begin{equation}
     s_{clk}(t)=cos(2\pi f_0 t + \frac{\Delta f}{f_m}sin(2\pi f_m t))
\end{equation}
where $f_0$ is the clock frequency, $f_m$ is modulating frequency, and $\Delta f$ is the peak frequency deviation. As a result, the frequency-domain clock signals can be derived as follows:
\begin{equation}
\begin{split}
\left\| \mathrm{F}(f)\right\| &= 
\left\| \sum_{n} J_n\left(\frac{\Delta f}{f_m}\right) \right. \\
&\quad \left. \times \big( \delta(f - f_0 + n f_m) 
- \delta(f - f_0 - n f_m) \big) \right\|
\end{split}
\end{equation}
where $J_n(\cdot)$ is the Bessel function and $\delta(\cdot)$ is the Dirac delta function. For the sake of simplicity, we can rewrite the spectrum expression in the above as follows: 
\begin{equation}
    s_{clk}(t)=\sum_{n=1}^{N}A_{clk}(n)sin(2\pi f_{clk}t)
\end{equation}
where $A_{em}(n)$ represents the amplitude of the  spikes and $f_{clk}=f_0 -nf_m$, or $f_{clk}=f_0+nf_m$. Now, let's consider the computational activity, such as a series of periodic memory accesses or switching behaviors, which can introduce the square waves at the clock edges. This square wave can be expressed with the Fourier transform as follows:
\begin{equation}
    s_{sq}(t)=\frac{2A_{sq}}{\pi} \sum_{m}\frac{cos((2m-1)2\pi f_{sq}t)}{2m-1}
\end{equation}
where $f_{sq}$ and $A_{sq}$ denote the frequency and amplitude of the square wave introduced by the computational activity. As a result, this computational activity introduces the amplitude modulation to the clock signals, which is equivalent to frequency mixing. Therefore, the emanations are produced as:
\begin{equation}
    s_{em}(t)=s_{clk}(t)s_{sq}(t)=\sum_{n}\sum_{m}\frac{A_{clk}(n)A_{sq}}{(2m-1)\pi}s_{p}(n,m,t)
\end{equation}
where $s_{p}(n,m,t)=sin(2\pi(f_{clk}+(2m-1)f{sq})t)+sin(2\pi(f_{clk}-(2m-1)f{sq})t)$ indicates clock-introduced spikes are modulated by the computational activity-introduced spikes. 

\begin{figure}
    \centering
    \includegraphics[width=\linewidth]{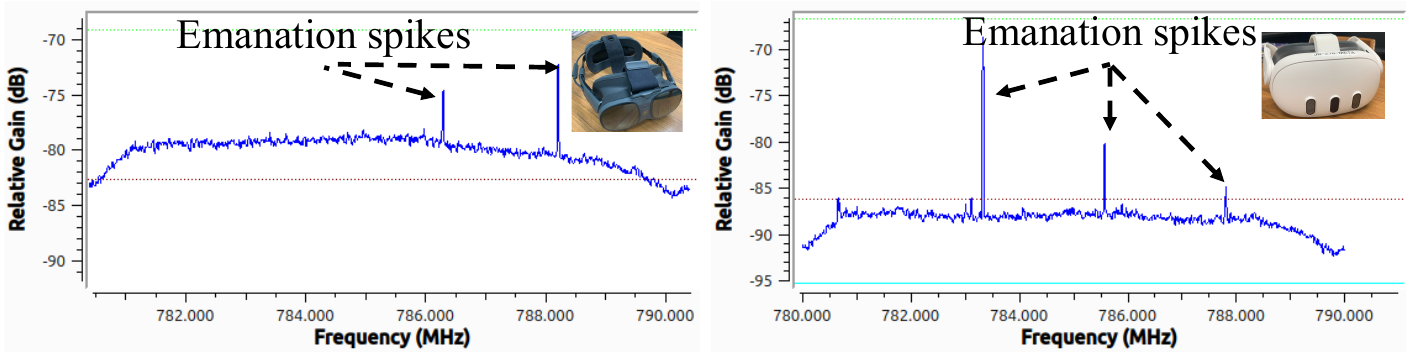}
    \caption{Frequency-domain representation of emanations from HTC VIVE XR Elite and Meta Quest 3.}
    \label{fig:em:example}
\end{figure}

\noindent \textbf{EM characterizations.} The emanations are clock signals that are modulated in amplitude by the computational activities or switching behaviors on the hardware devices. The clock signals are coupled with the switching behaviors or computational activities through different hardware components such as capacitors, resistors, and diodes. As a result, the emanations in the time domain exhibit the squared waves, where the periodicity is the clock frequency. In the frequency domain, the emanations exhibiting the harmonics can spread across a wide frequency band with multiples of emanation spikes, consisting of a fundamental harmonic and multiples of harmonics as shown in Fig.~\ref{fig:em:example}. The distance between the adjacent harmonics indicates the clock frequency. 

\noindent \textbf{EM security and privacy.} As we can see, the spreading emanation spectrum consists of a fundamental frequency spike and a series of multiple frequency spikes, which are determined by the computational activities at the emanation source (e.g., VR headset). These emanations of the VR headset can carry important information about the VR users, which can be carefully characterized through signal processing techniques (e.g., FFT or STFT). Then, we can leverage the machine learning models to reveal the hidden privacy information from the emanations, resulting in a great privacy threat.  However, since the energy of the emanations is spread across a wide spectrum, it is difficult to detect all the emanation harmonics emitted from different emanation sources like the prior works~\cite{jin2021periscope,vuagnoux2009compromising,wang2011analysis, sun2025revealing}. Moreover, the ambient wireless communication signals can interfere with the emanations, which need to be suppressed for accurate harmonic extraction. The interleaved emanations from the different sensors in the VR headset make the privacy attack more challenging. 

\begin{figure}
\centering
    \includegraphics[width=\linewidth]{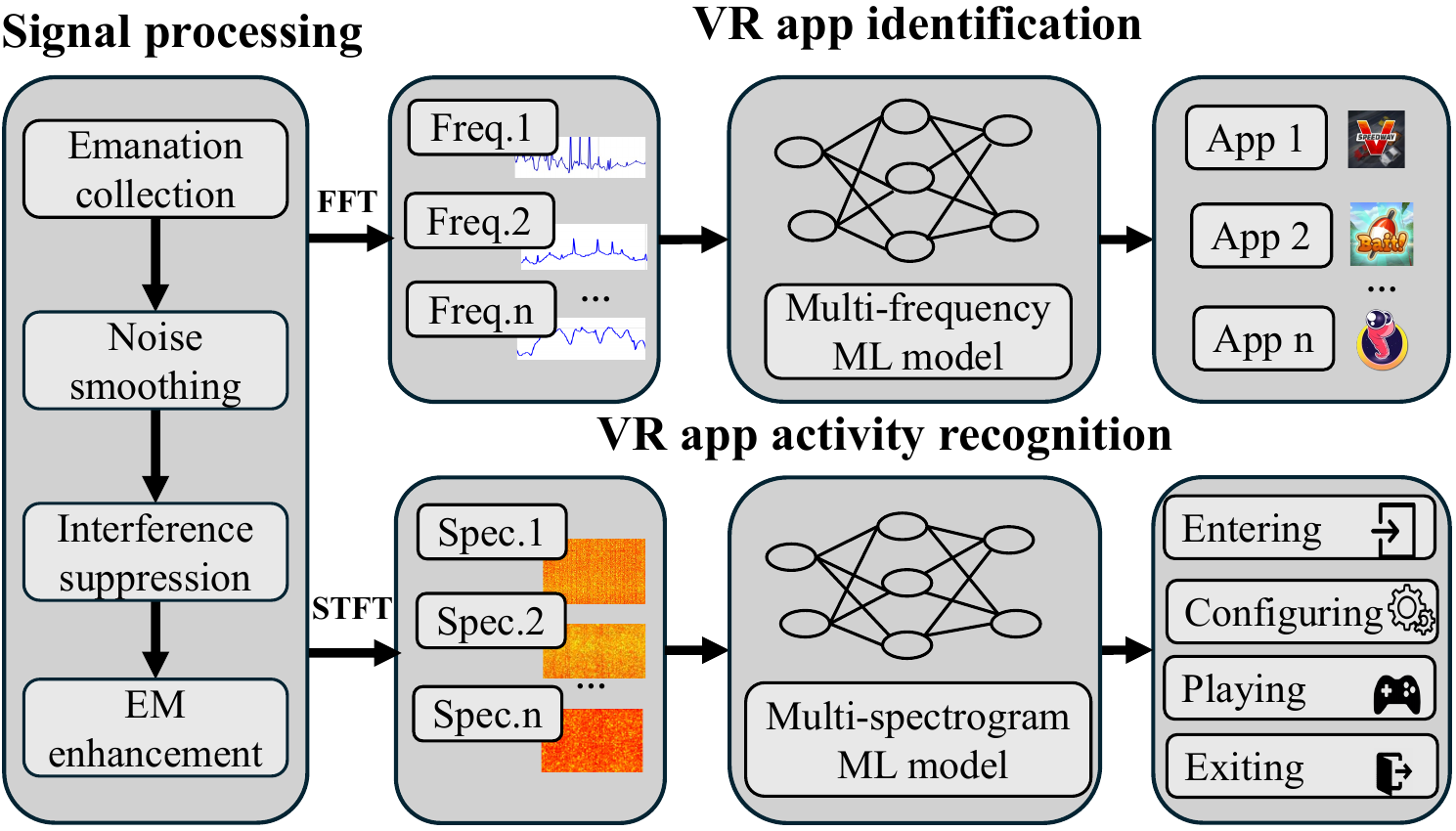}
    \caption{The workflow of \sysname consists of the emanation signal processing module, VR app identification module, and VR app activity recognition module.}
    \label{fig:work:pipeline}
\end{figure}

\section{Threat Model}
\label{sec:threat}

\noindent \textbf{Attack settings.} Our attack leverages the wireless signal receiver (e.g., software-defined radio USRP N210, PlutoSDR, LimeSDR, etc.) to passively sniff the electromagnetic emanations emitted from the VR headset. The VR user can wear the VR headset or the head-mounted display (HMD) to enjoy the embodied and immersive experience. We assume that the attacker can deploy a radio-frequency sniffer close to the victim for emanation eavesdropping. For example, the VR user plays the VR games in a public space (e.g., cafeteria, transportation) while the attacker is eavesdropping next to the victim~\cite{hayashi2014threat}. Similar attack scenarios were validated in emanation-based attacks~\cite{luo2024eavesdropping,choi2020tempest,hayashi2014threat, chen2024eavesdropping} for other purposes.

\noindent \textbf{Adversarial model.} We make the following assumptions about the adversary. 
\begin{itemize}[leftmargin=*]
    \item The adversary does not have physical or remote access to the victim's VR platform. However, the adversary can deploy the eavesdropping radio to sniff the electromagnetic emanations emitted from the VR headset.  
    \item The adversary only performs passive emanation eavesdropping using the RF radio instrumented with the directional antenna, which connects with the laptop to run the algorithms for VR app identification and activity recognition. 
    \item The goal of the attacker is to predict the VR app's identities (e.g., Aim app and Vspeedway app) and activities (e.g., entering or configuring the VR app), which can be further abused for app recommendation and the VR user's personality inference. 
    \item Similar to the attack proposed in~\cite{ni2023eavesdropping}, we assume the adversary mainly focuses on a set of VR apps and app activities. The inference of the VR user's personality and biometric behavior using eavesdropped VR app identities and activities is beyond the scope of this paper. 
\end{itemize}

\section{System Overview}
\label{sec:overview}

Fig.~\ref{fig:work:pipeline} shows the workflow of our eavesdropping system consisting of the signal processing module, app identification module, and app activity recognition module.
\begin{itemize}[leftmargin=*]
    \item \textbf{Signal processing module.} After the emanation signals are eavesdropped across a wide frequency band, we first smooth the noise floor with a mean filter to remove the frequency-dependent noise variation across the frequency bands. Then, we suppress the ambient interferences, including the ambient emanations and wireless communication signals, through spectrum subtraction. We further boost the emanation strength through non-coherent averaging of the eavesdropped signals over time. 
    \item \textbf{VR app identification module.} To identify the VR app, we design a fine-tuned pre-trained multi-frequency ML model to infer the VR app identities based on frequency-domain characteristics of the emanations using FFT. 
    \item \textbf{VR app activity recognition module.} To recognize the VR app activities, we design a fine-tuned pre-trained multi-spectrogram ML model to infer the app activities based on the spectrogram of emanations derived with STFT that can characterize over-time properties of the frequency-domain emanations. 
\end{itemize}

In what follows, we first present the emanation signal processing, including the experimental evaluation-based noise floor smoothing, ambient wireless interference suppression, and emanation strength enhancement. Then, we establish the relationship between the characterization of emanations and the app identities or activities through machine learning models. 

\section{System Design}
\label{sec:attack:design}

\begin{figure}
\centering
\captionsetup{width=0.23\textwidth}
\begin{minipage}{0.25\textwidth}
\centering
    \includegraphics[width=0.8\linewidth]{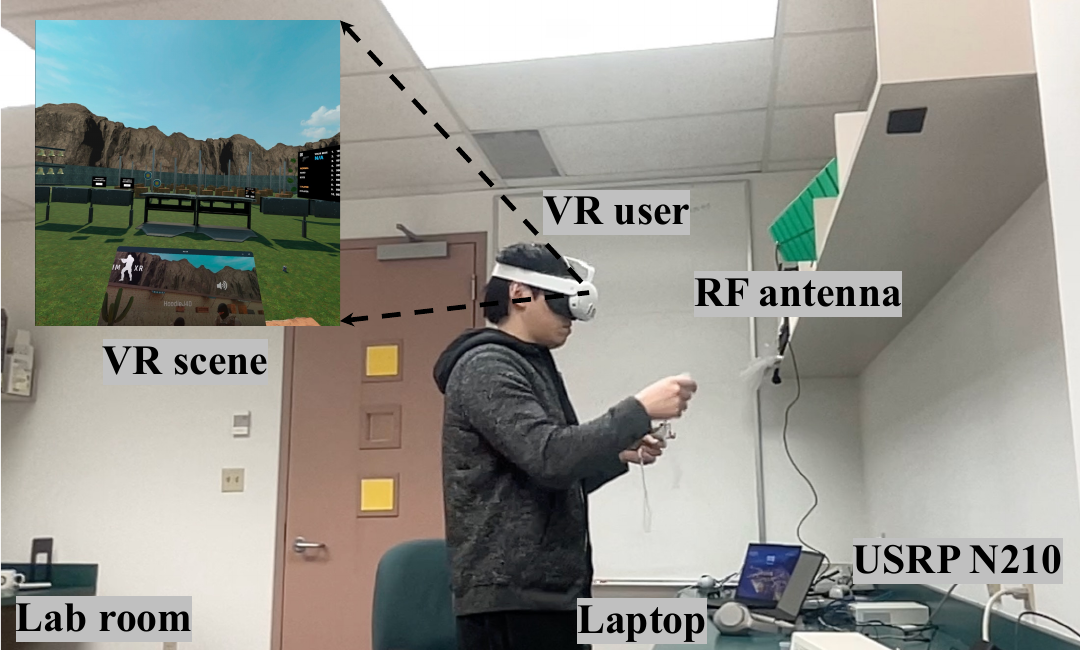}
    \caption{Experimental setup in the lab room, where the VR user wears the Meta Quest 3, and the USRP N210 connects with the directional antenna for emanation sniffing.}
    \label{fig:feasibility:setup}
\end{minipage}%
\begin{minipage}{0.25\textwidth}
  \centering
    \includegraphics[width=0.86\linewidth]{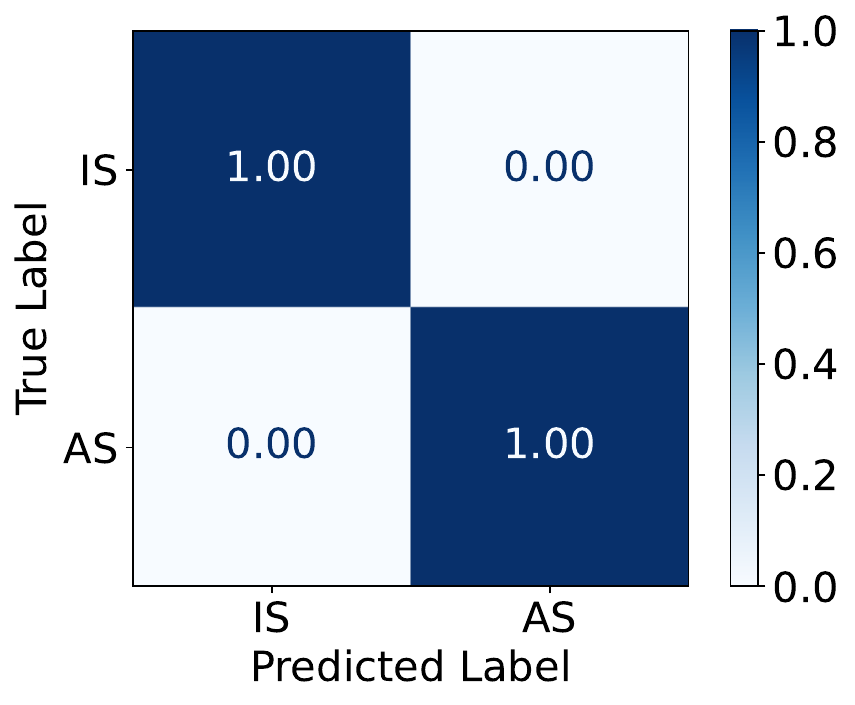}
    \caption{Confusion matrix of active state (AS) and idle state (IS) classification for VR headset.}
    \label{fig:cm:state:detection}
\end{minipage}
\end{figure}
\subsection{Emanation Signal Processing}
\label{subsec:signal:processing}

The emanations from the VR headset exhibit amplitude-modulated clock signals in the time domain and spread across a wide frequency band in the frequency domain. Our goal is to accurately extract and characterize the frequency-domain emanation signals for VR app identification and activity recognition. Our attack design options are demonstrated through the experimental measurements.

\noindent \textbf{Experimental settings.} To do so, we illustrate the experimental setup shown in Fig.~\ref{fig:feasibility:setup}. The VR user is wearing the headset of Meta Quest 3 or HTC VIVE XR Elite for an embodied and immersive experience. At the same time, we use the USRP N210 instrumented with the directional antenna and UBX daughterboard to sniff emanations. The USRP N210 connects to the laptop of the Lenovo ThinkPad for IQ sample processing. More details about our attack system implementation can be found in Section~\ref{sec:imp:eva}.  The adversary can detect the VR app-running state through the variation of the emanations introduced by the VR headset, as the emanations are mainly affected by the computational activities on the VR headset. As a demonstration, we leverage the emanation source state detection technique from IoTProsector~\cite{sun2023feasibility} to sense the VR app state, which can be used to guide the adversary for VR app-related emanation sensing. Fig.~\ref{fig:cm:state:detection} shows the confusion matrix for the VR app state detection, where the binary classification achieves $100\%$ accuracy. Hence, we accurately extract the VR app-related emanations for app identification and activity recognition. From this experimental setup, we explore the impact of noise and ambient wireless interferences on the emanations. Furthermore, we notice that the emanations should be strengthened for detection.

\noindent \textbf{Noise floor smoothing.}  Since the frequency-domain emanations are spreading across a wide frequency band, the varying noise floor across the spectrum needs to be smoothed for accurate emanation characterization. As such, we need to scan a wide spectrum for emanation spike characterization. To do so, we first scan the frequency band below 1 GHz, using USRP N210 as a receiver instrumented with the LP0410 PCB antenna~\cite{lp0410antenna} with a sampling rate of 25 MHz and a bandwidth of 10 MHz. Then, we concatenate all the 10 MHz frequency bands for emanation characterization. We apply the movmedian filter~\cite{movmedian} for noise removal. We further smooth the noise floor across a wide frequency band. After the noise removal, the noise floor across the wide frequency band becomes flat and smooth, which can advance the frequency-domain emanation characterization.

\begin{figure}
\centering
    \includegraphics[width=0.86\linewidth]{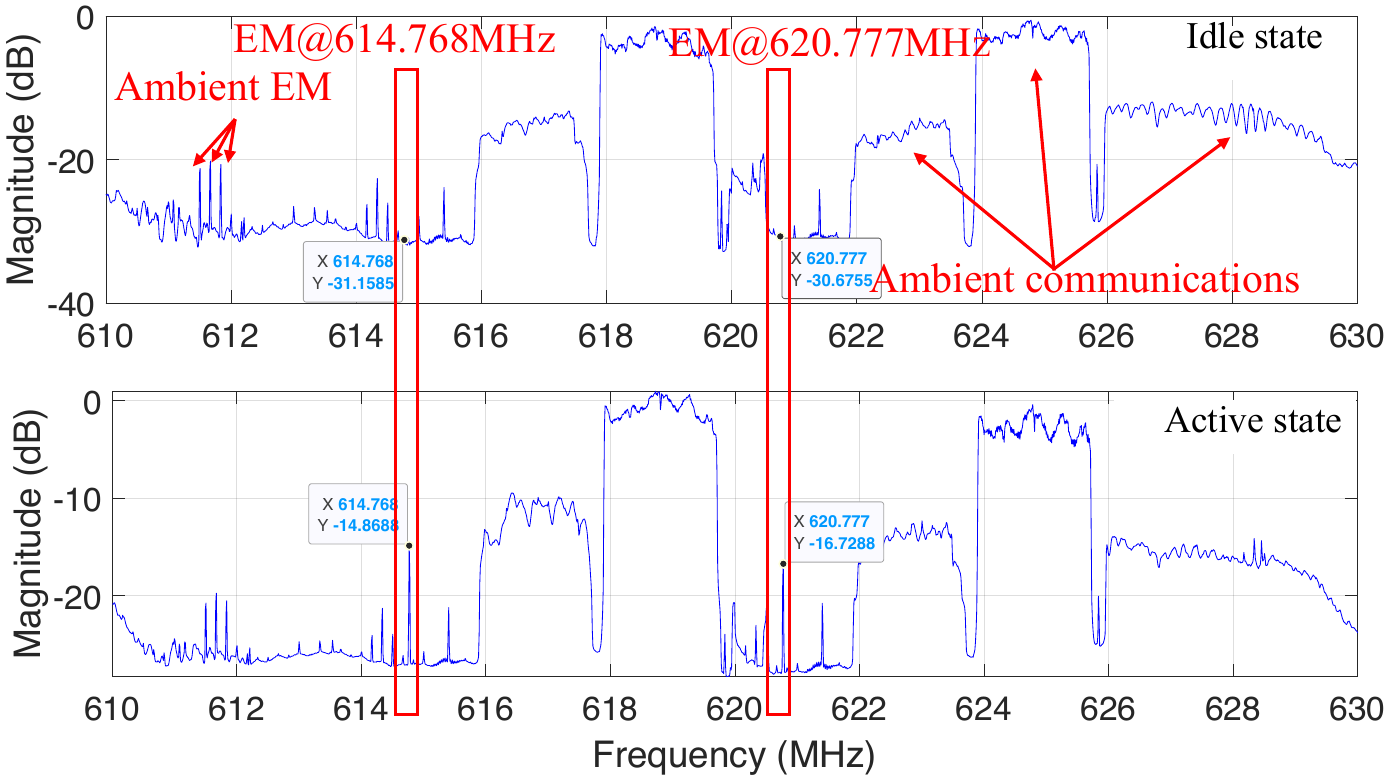}
    \caption{FFT of the eavesdropped wireless signals during active and idle states of VR application usage, showing the emanations from the VR headset, ambient emanations, and wireless communication signals.}
    \label{fig:em:grandtour}
\end{figure}

\noindent \textbf{Ambient wireless interference suppression.} As shown in Fig.~\ref{fig:em:grandtour}, the sniffed wireless signals at the eavesdropper include the ambient emanations from the electronic devices (e.g., monitors), ambient wireless communication signals (e.g., cellular signals), and emanations from the VR headset. As we can see, the spectrum is primarily dominated by ambient wireless communication signals, which need to be mitigated for accurate VR headset-introduced emanation characterization. To this end, we propose to eliminate these ambient artifacts through subtraction. Specifically, we first characterize the ambient wireless communication environment. We assume that these ambient artifacts do not change abruptly, which has been experimentally demonstrated in the prior work~\cite{sun2025revealing}. As such, we can eliminate these artifacts by using a sliding window-based spectrum subtraction over the sniffed wireless signals.

\begin{figure}
\centering
    \includegraphics[width=0.8\linewidth]{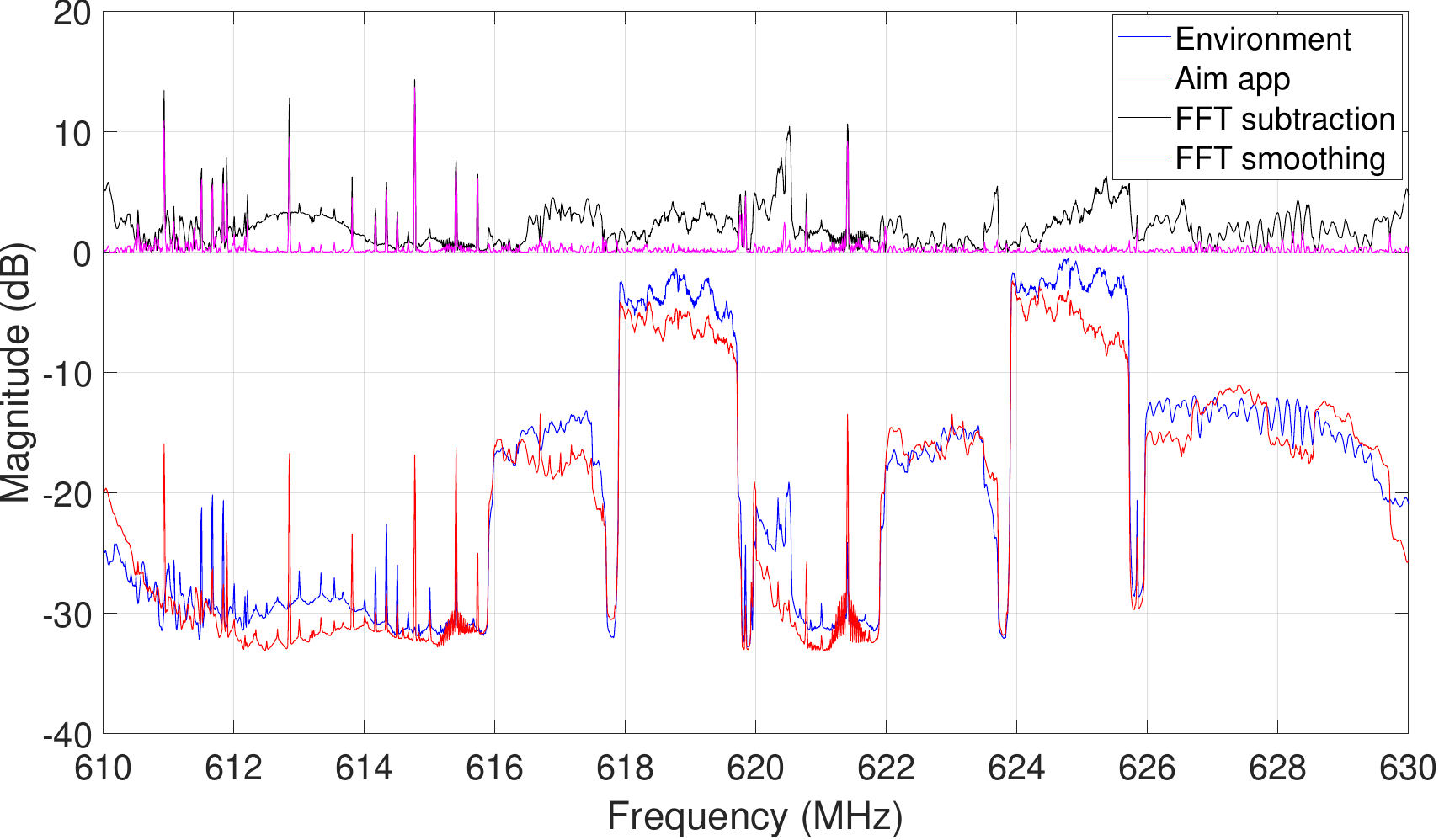}
    \caption{FFT of the emanations in the physical wireless environment during the active and idle states of application usage, and the FFT of the emanations after ambient wireless signals subtraction and noise floor smoothing.}
    \label{fig:smooth:subtraction}
\end{figure}

Fig.~\ref{fig:smooth:subtraction} shows the efficiency of noise floor smoothing with the noise removal filter and interference suppression through subtraction. We first show the FFT of the received wireless signals indicated by the blue line in the figure when the VR headset is in the idle state. Then, we show the FFT of the received wireless signals indicated by the red line in the figure when the VR headset is in the active state. As we can see, most of the spectrum (e.g., the ambient wireless communication signals) is overlapped due to the quasi-stable wireless environment, while the emanation spikes are outstanding when the VR headset is in the active state. After we do subtraction between them, as indicated in the black line in the figure, the ambient wireless communication signals are suppressed. To further remove the variation of the noise floor, we apply the movmedian filter~\cite{movmedian} on the subtracted spectrum, as indicated in the pink line. As such, the noise floor becomes flat and smooth across the frequency bands. Moreover, we can see the outstanding emanations spikes in the FFT result, which can be leveraged for app identification and activity recognition.

\begin{figure}
  \centering
\includegraphics[width=0.55\linewidth]{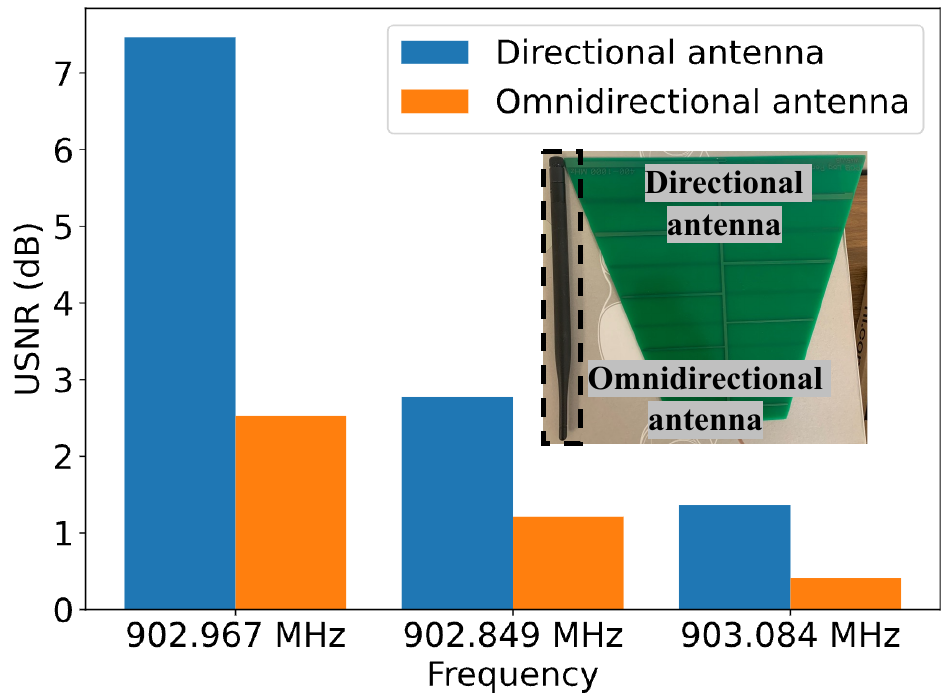}
    \caption{Impact of the different antenna gains on USNR of the eavesdropped emanation.}
    \label{fig:antenna:gain}
\end{figure}

\begin{figure}
\centering
    \includegraphics[width=\linewidth]{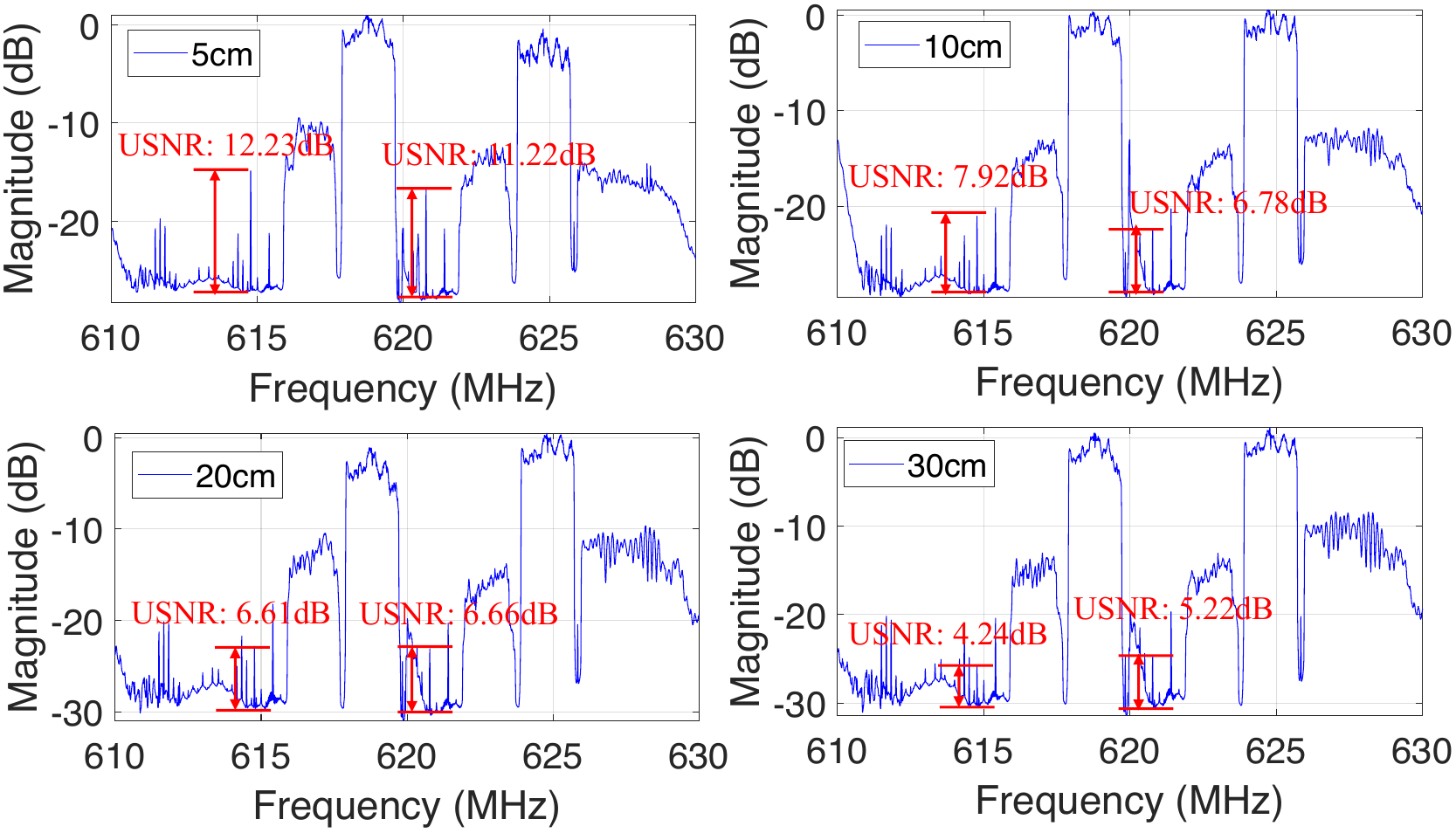}
    \caption{USNR of the emanations from the VR headset over different distances.}
    \label{fig:em:distance}
\end{figure}

\noindent \textbf{EM strength enhancement.} Since emanations are unintentionally emitted from the VR headset, which is naturally weak in strength. The weak emanation signals restrict the emanation detection and characterization. To enhance the emanation strength received at the eavesdropper, the straightforward idea is to use a high-gain directional antenna or amplifier. To demonstrate this, we compare the unintentional signal-to-noise ratio (USNR)~\cite{choi2020tempest} of the eavesdropped emanations when we use an omnidirectional antenna (i.e., VERT900 with 3dBi gain) and a directional antenna (i.e., LP0410 with 6dBi gain) instrumented on the USRP N210. Fig.~\ref{fig:antenna:gain} shows the USNR of emanation spikes using antennas with different gains. As we can see, the directional antenna with more power gains can eavesdrop on the emanations with higher strength than the omnidirectional antenna. This indicates that we can use a high-gain directional antenna for emanation eavesdropping. We can also use a power amplifier at the eavesdropper to enhance the sniffed emanation strength. Fig.~\ref{fig:em:distance} shows USNR over different distances between the attacker and VR user. As we can see, as the distance between the attacker and the VR headset becomes larger, the USNR becomes smaller due to the path loss. Therefore, it is important to boost the emanation signal strength for long-range emanation detection and characterization.

To further enhance the emanation strength and even bring up the emanations below the noise floor, we propose to boost the emanation signal strength by exploiting the emanation signal characteristics. Specifically, since the emanations are amplitude-modulated clock signals, they are represented as square waves in the time domain. However, the noise does not exhibit any specific pattern. So, we can use the non-coherent averaging to boost the emanation strength by taking the average of the over-time emanation signals, while the noise can be averaged out. As such, the emanations become outstanding in the spectrum.

\begin{figure}
\centering
\captionsetup{width=0.23\textwidth}
\begin{minipage}{0.25\textwidth}
\centering
    \includegraphics[width=\linewidth]{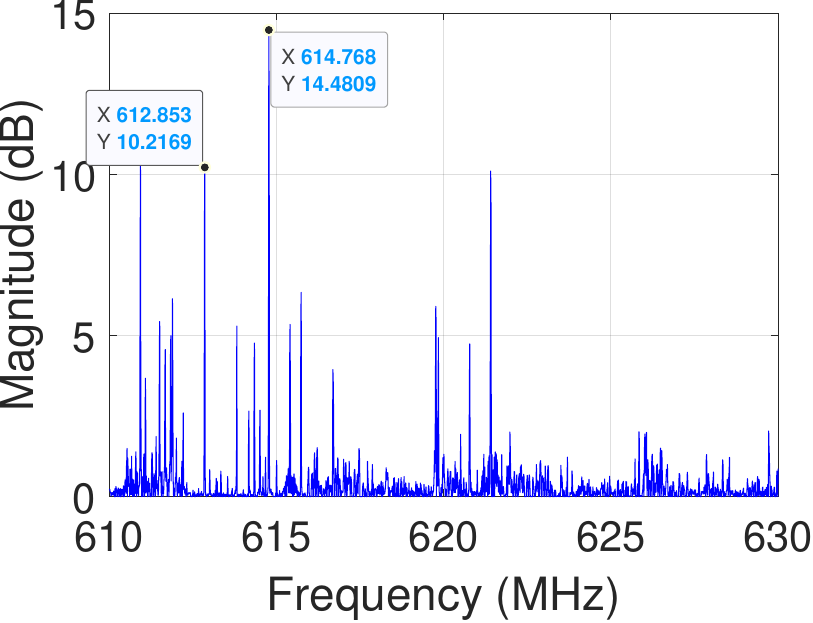}
    \caption{Averaging FFT on the emanations over 0.1s.}
    \label{fig:1s}
\end{minipage}%
\begin{minipage}{0.25\textwidth}
  \centering
   \includegraphics[width=\linewidth]{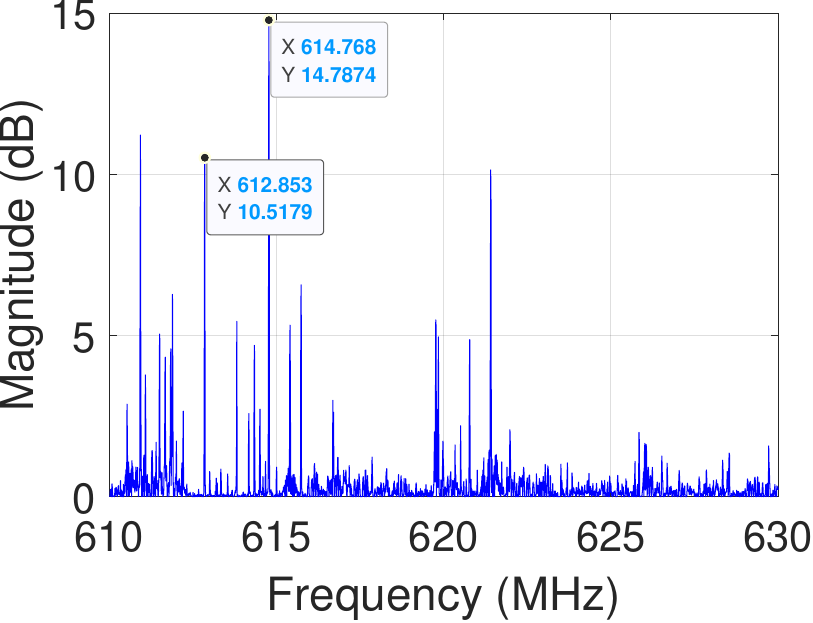}
    \caption{Averaging FFT on the emanations over 0.2s.}
    \label{fig:2s}
\end{minipage}
\end{figure}

Fig.~\ref{fig:1s} and Fig.~\ref{fig:2s} present the averaged FFT of the emanations emitted from the VR headset when we run the VR app. As we can see, when we do the averaging FFT on the eavesdropped emanations over 0.2s, the USNR of the emanation spike becomes larger in comparison to the averaging FFT on the eavesdropped emanations over 0.1s. Specifically, the emanation spike at the frequency of 614.768 MHz exhibits a USNR of 14.7874 dB when we use averaging FFT on the emanations over 0.2s, while it is 14.4809 dB when we use averaging FFT on the emanations over 0.1s. As such, we have the USNR gain of 0.2 ($=\frac{14.7874-14.4809}{14.4809}$)dB per second. So, averaging over a longer time duration could boost the emanation signal strength. For example, averaging over 10s could boost the emanation strength by 2 dB. This is because the emanations are square waves, while the noise does not show any specific pattern. As such, averaging the FFT could improve the emanation's USNR. 

\subsection{Machine Learning-based App Identities and Activities Inference}
\label{subsec:ml:inference}

To identify the VR apps and infer their activities, we propose to use machine learning based on the characterized emanations. Specifically, we can leverage the FFT of the emanations to infer the app identities.

\begin{figure}
\centering
\captionsetup{width=0.23\textwidth}
\begin{minipage}{0.25\textwidth}
\centering
    \includegraphics[width=\linewidth]{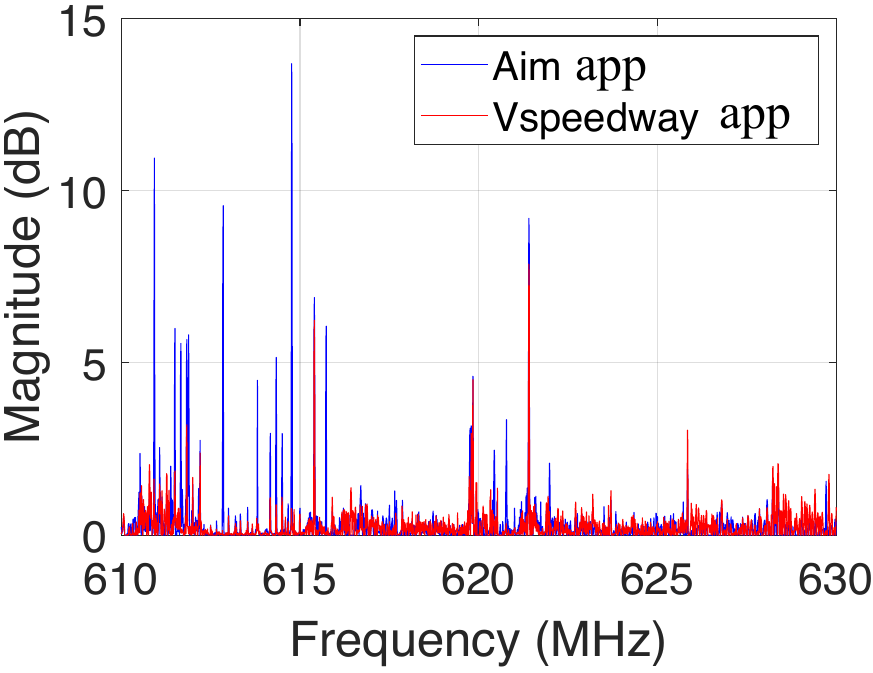}
    \caption{FFT of emanations from VR headset running different VR apps.}
    \label{fig:fft:aim:gaming}
\end{minipage}%
\begin{minipage}{0.25\textwidth}
  \centering
    \includegraphics[width=\linewidth]{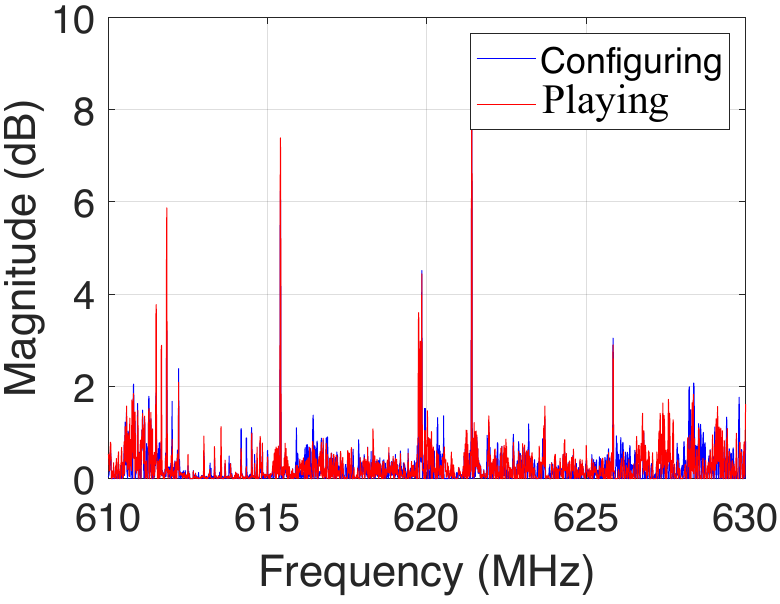}
    \caption{FFT of emanations from VR headset configuring or running the VR app.}
    \label{fig:fft:vspeedway:gaming:configuration}
\end{minipage}
\end{figure}

\noindent \textbf{App identification.} To demonstrate the feasibility of using FFT to discriminate different VR apps, we conduct an experiment to show the FFT results when we run different VR apps. As shown in Fig.~\ref{fig:fft:aim:gaming}, the frequency-domain emanations are different when we run different VR apps. This is because the emanations are amplitude-modulated clock signals, which are mainly affected by the computational activities on the VR headset. As a result, the frequency-domain emanations can carry VR app information, which can be used to infer VR app identities.

To accurately characterize the frequency-domain emanations, we design a neural network to classify the VR apps. Our multi-frequency neural network should characterize over-frequency emanation spikes. To this end, we propose to leverage the ResNet~\cite{targ2016resnet} with a fine-tuned convolutional layer that can adapt to the input of FFT across multiple frequency bands and further infer the VR app identities. This is because the pre-trained ResNet has been demonstrated to be efficient in characterizing the wireless environment. As such, we can infer the VR app identity accurately. Specifically, the input of the ResNet is the frequency-domain spikes, which have already eliminated the effect of the physical wireless environment through subtraction. The output can be used to classify app identities. 

\begin{figure}
    \centering
    \includegraphics[width=0.86\linewidth]{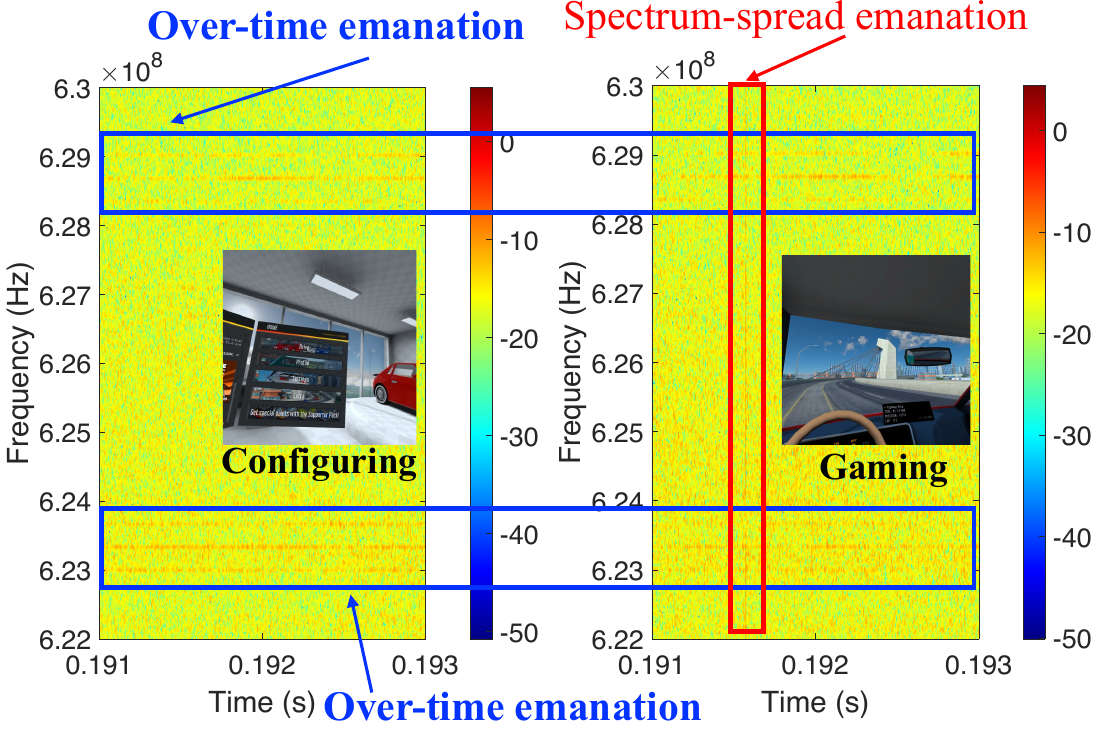}
    \caption{Spectrogram of emanations when we configure the app (left figure) and run the app (right figure). The blue rectangles show the different over-time emanation patterns for gaming and configuring. The red rectangle shows the spread-spectrum emanation only when the VR user runs the app.}
    \label{fig:em:grandtour:spectrum}
\end{figure}

\noindent \textbf{App activity recognition.} To infer the app activities, can we still use the FFT of emanations to discriminate the VR app's activities? To answer this question, we conduct an experiment to show the FFT results when we configure and run a VR app. As shown in Fig.~\ref{fig:fft:vspeedway:gaming:configuration}, the emanation spikes overlap with each other. This is because the same app is configured or operated on the VR headset. Therefore, it is not possible to differentiate the VR app's activities based on the frequency-domain emanations alone. However, we find that different activities running on the VR headset can introduce over-time characteristics to the frequency-domain emanations. To characterize the over-time properties of the emanations, we exploit the spectrogram that can show the over-time and over-frequency features of the emanations. As we can see in Fig.~\ref{fig:em:grandtour:spectrum}, the left and right spectrograms present the emanations in the time and frequency domain when we run and configure a VR app. As we can see, the red rectangle shows the spectrum-spread emanations across the frequency, which are presented only when the VR user runs the app. Moreover, the blue rectangles show the over-time emanations for the in-running and in-configuration VR app, which exhibit different over-time patterns. This is because the emanations are determined by the computational activities on the VR headset, resulting in differentiated over-time emanations. Therefore, we can leverage the spectrogram with time and frequency-domain features of the emanations derived with the short-time Fourier transform (STFT) to infer the VR app activities. Specifically, we leverage the multi-spectrogram machine learning model using ResNet architecture with a fine-tuned convolutional layer that can take the spectrogram as input to recognize the VR app activity, including entering, configuring, running, and exiting. 

Since we leverage frequency-domain emanations and the over-time frequency-domain emanations to infer app identity and activity, we cannot simply use one machine learning model for these two tasks through multi-task learning. Moreover, the app activity recognition is based on the over-time frequency characterization of the emanations. To this end, after the attacker eavesdrops on the IQ samples, the frequency-domain emanations are used to predict the app identities, and the over-time frequency-domain emanations are used to predict the app activities. These two tasks have been performed in parallel since they are independent of each other, which can potentially accelerate the attack process. Next, we illustrate the implementation and evaluation details.

\section{Implementation and Evaluation}
\label{sec:imp:eva}

\begin{figure}
    \centering
\includegraphics[width=0.68\linewidth]{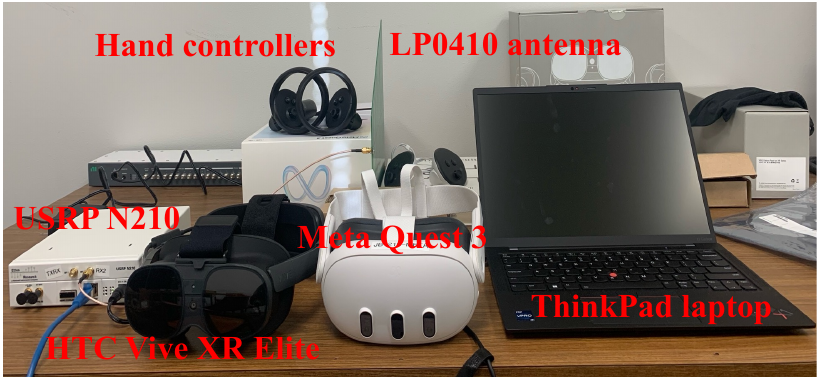}
    \caption{Experimental devices used for system performance evaluation.}
    \label{fig:experimental:hardwares}
\end{figure}

\noindent \textbf{Hardware and software.} To eavesdrop on the emanations from the VR headset (e.g., Meta Quest~\cite{quest} or HTC VIVE XR Elite~\cite{htv}), as shown in Fig.~\ref{fig:experimental:hardwares}, we use a UBX40 daughterboard-enabled USRP N210~\cite{usrp} as the software-defined radio (SDR) instrumented with the directional antenna LP0410~\cite{lp0410antenna}, which is also compared with the omnidirectional antenna VERT900~\cite{vert900} on USNR~\cite{choi2020tempest} measurements. The USRP N210 connects to the ThinkPad X1 Carbon Gen11 laptop with an Intel i7 CPU running Ubuntu 20.04 OS. The sniffed IQ samples are streamed to this laptop for further analysis. Specifically, USRP N210 streams the sniffed IQ samples to the laptop, which runs our signal processing algorithms in MATLAB for emanation characterization in the frequency domain. Our fine-tuned machine learning models are well-trained and implemented with Pytorch on the server with an NVIDIA A6000 Ada GPU for app identification and activity recognition. Our Pytorch code consists of FFT and Hamming window-based STFT on the IQ samples for app identification and activity recognition, respectively. We use a fine-tuned pre-trained ResNet18 as our foundational neural network model with a fine-tuned convolutional layer and output layer. 

\begin{figure}
    \centering
\includegraphics[width=\linewidth]{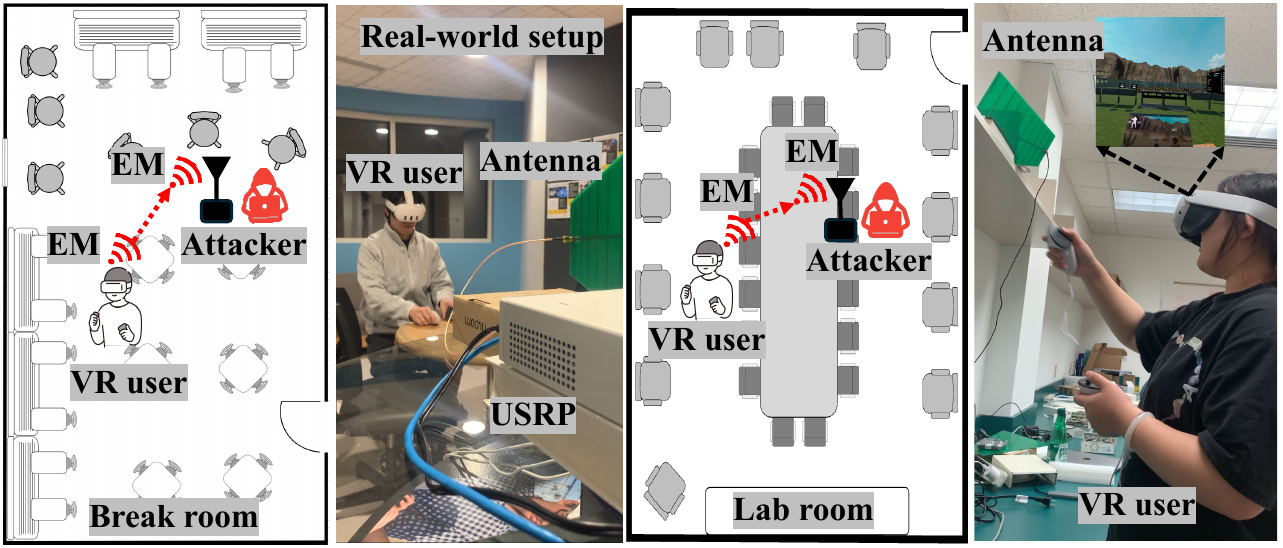}
    \caption{Experimental settings in the break room and lab room of the departmental building for system performance evaluation.}
    \label{fig:experimental:setup}
\end{figure}

\noindent \textbf{Experimental settings.} Our experimental setup consists of the VR user (i.e., victim) wearing the VR headset and an attacker eavesdropping on emanations. We consider a list of fifteen VR apps (i.e., Aim, Bait, Epic, Slupies, Vspeedway, Beast, Duck, Stable, Master, Tennis, Cosmicflow, Openbrush, Hyperdash, Maestro, and Conjure cards apps) and four app-specific user activities (i.e., entering, configuring, running, and exiting operation) that are also considered in prior work~\cite{ni2023eavesdropping}. The VR users can run apps in the break room and lab room of the departmental building with rich ambient wireless signals, while the attacker eavesdrops on the emanations from the VR headset at any random locations that are 1 to 2 meters away from the VR user, as shown in Fig.~\ref{fig:experimental:setup}. Our study has received IRB approval from our institute. We scan the spectrum below 1 GHz with a sampling rate of 25 MHz and a bandwidth of 10 MHz. We experimentally notice that the emanations are mainly emitted from the frequency band between 580 MHz and 630 MHz. As such, we mainly eavesdrop on this frequency band. Eavesdropping on more frequency bands could potentially improve the performance of emanation-based app identification and activity recognition. We explore this impact in our experimental results section. We collect 7.5G of emanation IQ samples for app identification and 42.5G for app activity recognition. By default, we report the system performance evaluation with Meta Quest 3, and the collected time-domain emanations can be divided into chunks with 500K IQ samples for FFT and STFT. Then, we split the data set into a training set with a size of $70\%$, a validation set with a size of $15\%$, and a test set with a size of $15\%$. The best well-trained model is used for prediction.

\noindent \textbf{Evaluation metrics.} To evaluate the performance of the end-to-end system, we report the confusion matrix and accuracy for VR app identification and activity recognition. We evaluate different factors that can affect system performance, such as the impact of eavesdropper-victim distance, the emanation time duration, the number of frequency bands, and different VR headsets. We also use USNR~\cite{choi2020tempest} as a metric to show the emanation strength. To demonstrate the efficiency of our proposed fine-tuned ResNet, we also evaluate the performance of other deep neural network models (e.g., LSTM and transformers) on the VR app identification and activity recognition. We also propose a countermeasure for this eavesdropping attack and further show its potential through simulated emanations.

\section{Experimental Results}
\label{sec:exp}

\subsection{System-Level Evaluation}
\label{subsec:end:to:end}

\noindent \textbf{Method.} To evaluate the end-to-end system performance, we report the performance of the VR app identification and activity recognition. Specifically, we report the system performance on app identification and activity recognition through the accuracy and confusion matrix. Our emanation measurements are sniffed in the beakroom and lab room of the departmental building, as described in the implementation and evaluation section. The fine-tuned ResNet model is well-trained on the training dataset, and the best model is selected based on the validation set for prediction on the test set. The final performance is reported based on the test dataset. By default, we use emanations spreading across five bands for the evaluation.

\begin{figure}
\centering
\captionsetup{width=0.23\textwidth}
\begin{minipage}{0.25\textwidth}
\centering
    \includegraphics[width=\linewidth]{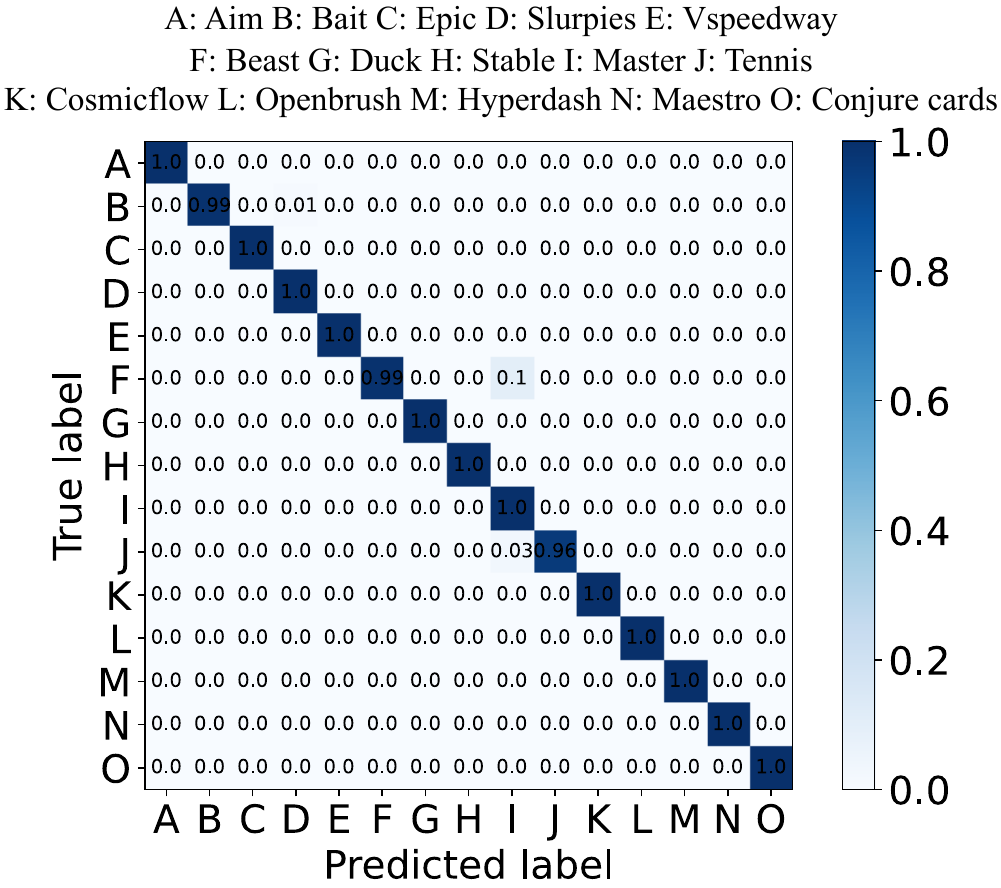}
    \caption{Confusion matrix of app identification.}
    \label{fig:cm:overall:types}
\end{minipage}%
\begin{minipage}{0.25\textwidth}
  \centering
    \includegraphics[width=\linewidth]{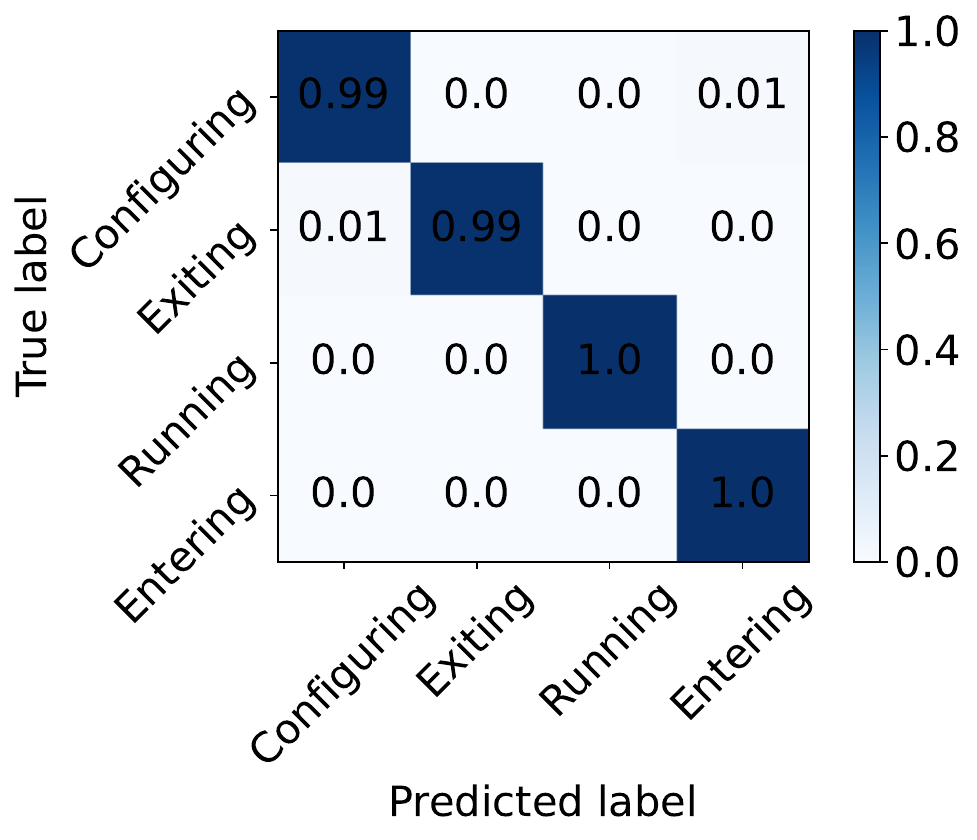}
    \caption{Confusion matrix of app activity recognition.}
    \label{fig:cm:overall:activity}
\end{minipage}
\end{figure}

\noindent \textbf{Result.}  Fig.~\ref{fig:cm:overall:types} and Fig.~\ref{fig:cm:overall:activity} show the confusion matrix of app identification and activity recognition. As we can see, these experimental results show the efficiency of our system in eavesdropping on VR app identities and activities via emanations. The high accuracy of the system performance is not due to the model being over-fitted, as we employ strategies (e.g., separate test and validation datasets, training adjustments, and fine-tuning pre-trained ResNet) to avoid it. Moreover, we use a fine-tuned ResNet model that can accurately characterize our signal processing-extracted emanations. Even though we use different fine-tuned ResNet models for app identification and app activity recognition, they are different from the input (i.e., FFT v.s. STFT results). As such, our attack can be conducted independently. This indicates the efficiency of the attack in identifying VR apps and recognizing app activities.

\subsection{Case Study}
To demonstrate the effectiveness of a stealthy attack using our proposed \sysname system, we conduct a case study. Similar to the attack scenarios reported in~\cite{hayashi2014threat},  in a typical public space, such as a cafe, a library, or a railway station, when VR users use VR devices to play games or chat online, the attacker can hide the eavesdropping setup in the backpack to closely sniff the emanations from the VR headset for app identification and activity recognition. 

\begin{figure}
\centering
\captionsetup{width=0.23\textwidth}
\begin{minipage}{0.25\textwidth}
\centering
    \includegraphics[width=0.85\linewidth]{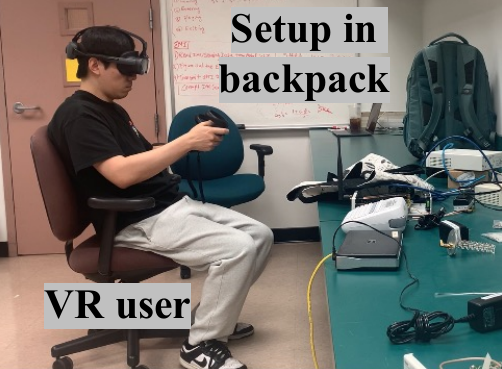}
    \caption{Concealed experimental setup in the case study.}
    \label{fig:case:study:setup}
\end{minipage}%
\begin{minipage}{0.25\textwidth}
  \centering
   \includegraphics[width=\linewidth]{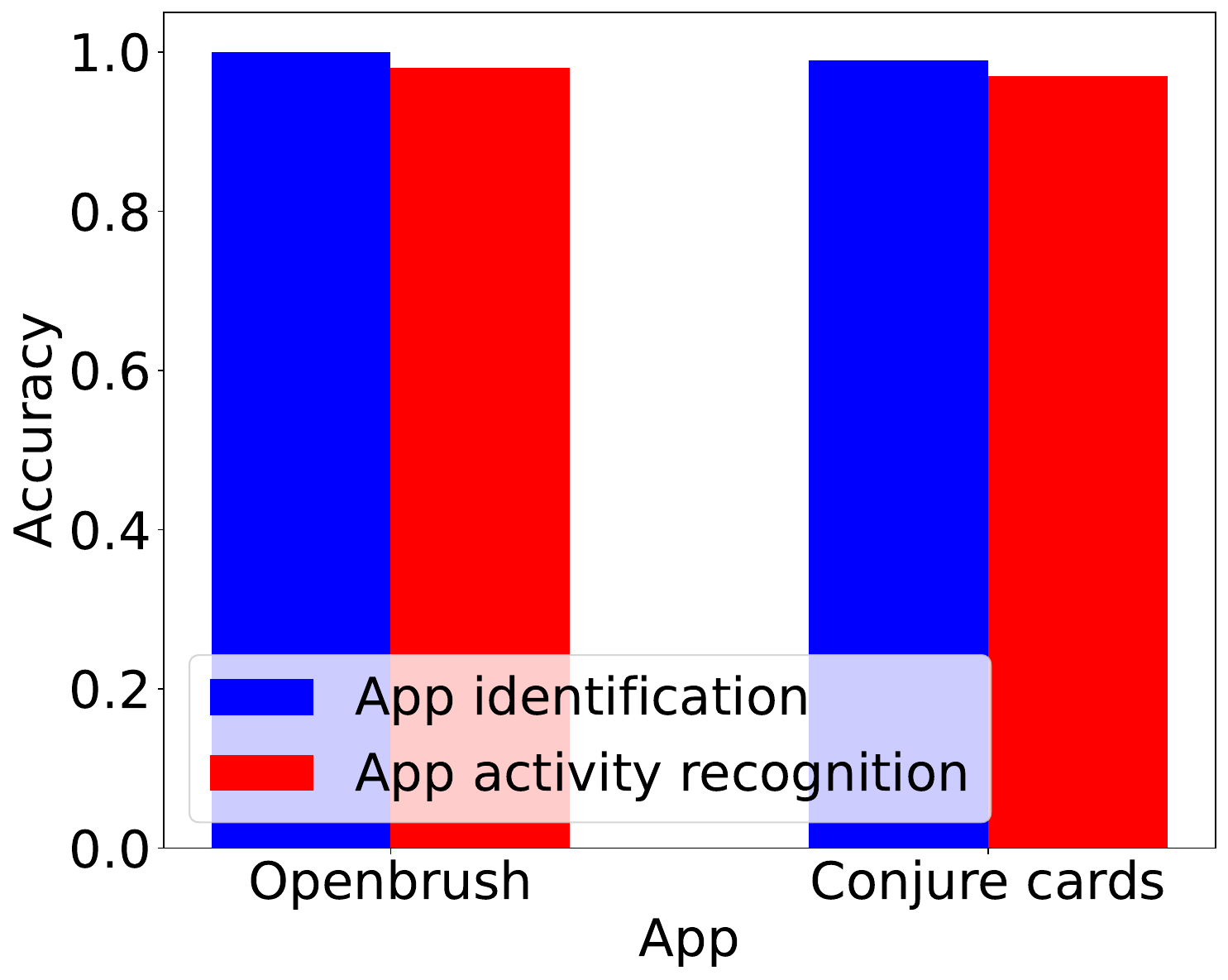}
    \caption{Performance of \sysname in the case study.}
    \label{fig:case:study:performance}
\end{minipage}
\end{figure}

\noindent \textbf{Method.} As an illustration, we can consider the scenario as shown in Fig.~\ref{fig:case:study:setup}, where the attack setup is hidden in the backpack to ensure the stealthiness and further eavesdrop on the emanation measurements from the VR headset and the VR user is running the VR app. The distance between the VR headset and the backpack is within 1 meter. This concealed setup is practical, which can be used in the public space (e.g., a cafe or a library) for eavesdropping similar to the attack scenario proposed  in~\cite{hayashi2014threat}. We use the well-trained model in the subsection~\ref{subsec:end:to:end} to predict the VR app identities and activities based on the emanation measurements collected with this concealed attack setup.  

\noindent \textbf{Result.} Fig.~\ref{fig:case:study:performance} shows the accuracy of app identification and activity recognition in the concealed setup. As we can see, the accuracy of app identification and activity recognition is around 0.99, which indicates the superiority of \sysname in the concealed or hidden setup. This is because \sysname is powered by the emanation enhancement techniques proposed in Section~\ref{sec:attack:design} and can be deployed closely to the VR users. 

\subsection{Microbenchmarks}

\subsubsection{Impact of VR headsets.} To see the impact of the VR headset on the system performance, we evaluate VR app identification and activity recognition across different VR headsets.

\noindent \textbf{Method.} To do so, we mainly use Meta Quest 3 and HTC Vive XR Elite to evaluate the impact of different VR headsets on VR app identification and activity recognition. Since the different brands of VR headsets do not share the same VR apps, we train the model on the emanation measurements collected and report the performance on the emanation measurements collected from different VR headsets separately. 

\begin{figure}
\centering
\captionsetup{width=0.23\textwidth}
\begin{minipage}{0.25\textwidth}
\centering
    \includegraphics[width=\linewidth]{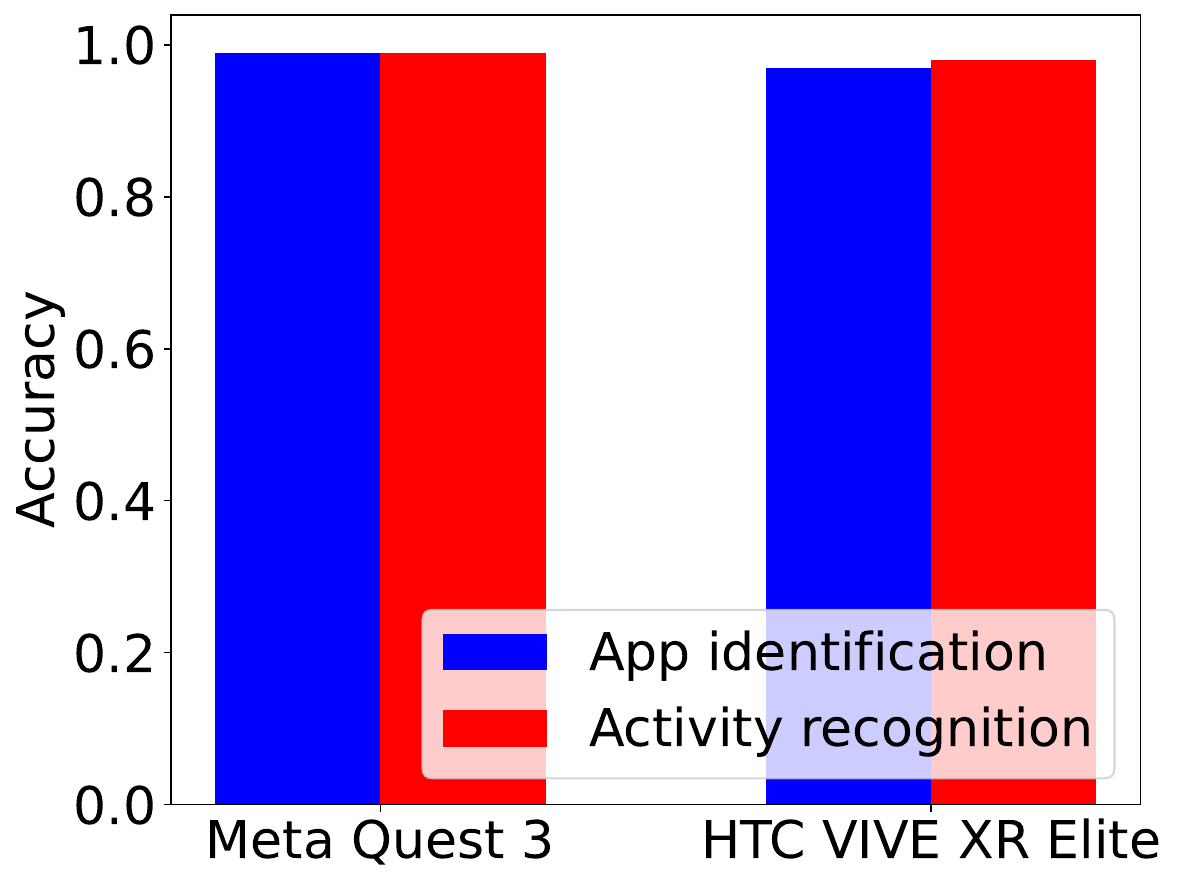}
    \caption{Accuracy of app identification and activity recognition over different VR headsets.}
    \label{fig:two:vr:acc}
\end{minipage}%
\begin{minipage}{0.25\textwidth}
  \centering
    \includegraphics[width=\linewidth]{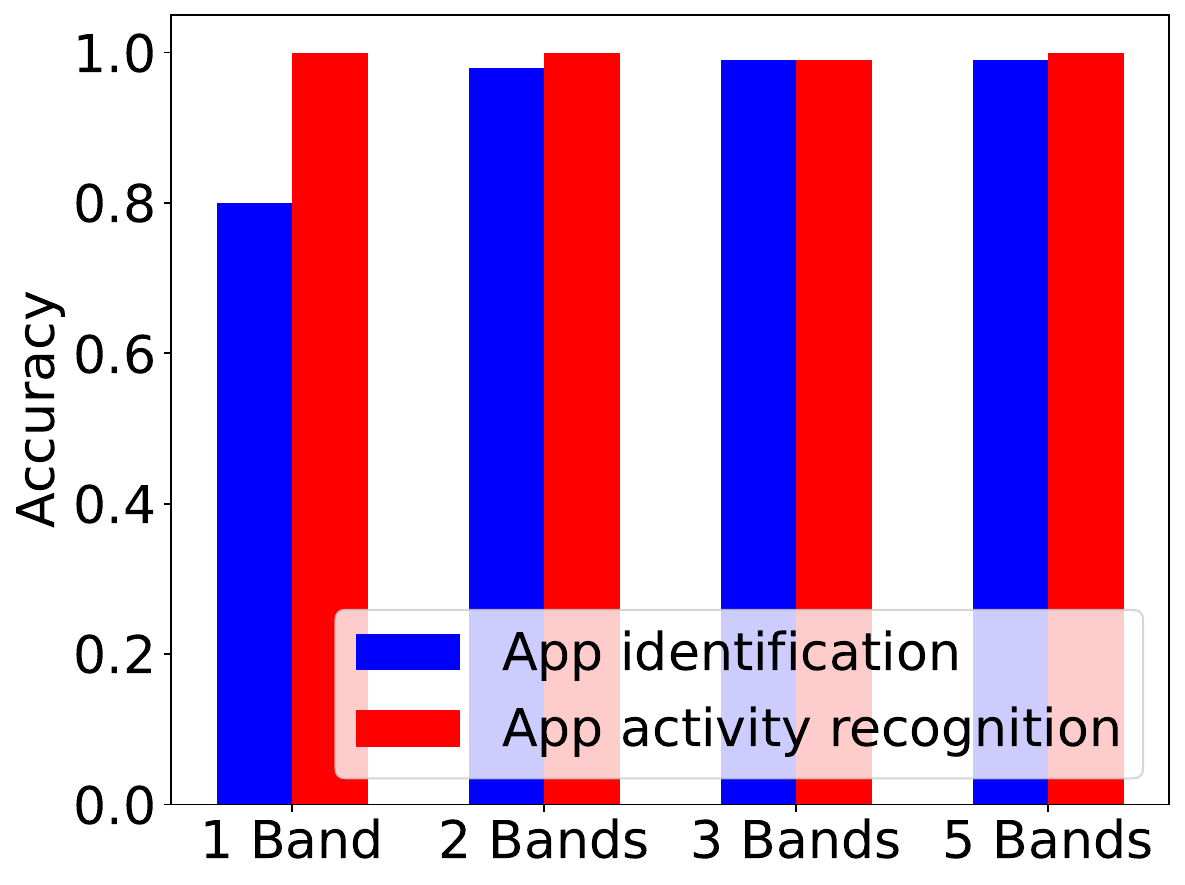}
    \caption{Accuracy of app identification and activity recognition across different numbers of bands.}
    \label{fig:app:accuracy:bands}
\end{minipage}
\end{figure}

\noindent \textbf{Result.} Fig.~\ref{fig:two:vr:acc} shows the accuracy of app identification and activity recognition with Meta Quest 3 and HTC VIVE XR Elite. As we can see, the accuracy of app identification and activity recognition over different VR headsets exhibits almost the same high values, as our subtraction method could not only suppress the ambient wireless interference but also eliminate the hardware-dependent artifacts. This demonstrates the efficiency of our proposed attack system over different VR headsets. 


\begin{table}
\centering
\caption{Performance comparison over different ML models.}
\begin{tabular}{c||c|c}
\hline
 & \multicolumn{2}{c}{\textbf{Accuracy}} \\ \hline
\textbf{ML models} & \textbf{App identification} & \textbf{Activity recognition}  \\ \hline
LSTM            & 0.73 & 0.71 \\ \hline
Transformer    & 0.76  & 0.75 \\ \hline
\textbf{Our model} & \textbf{0.99} & \textbf{0.99} \\ \hline  
\end{tabular}
\label{tb:ml:comparision}
\end{table}

\subsubsection{Impact of ML models.} Since the emanation measurements exhibit a specific pattern, we need to exploit the impact of the different machine learning models on characterizing the VR emanations.

\noindent \textbf{Method.} To demonstrate the efficiency of using fine-tuned ResNet for VR app identification and activity recognition, we compare it with the customized LSTM and transformer models that have already shown strong capability of characterizing the wireless spectrum.

\noindent \textbf{Result.} As shown in Table~\ref{tb:ml:comparision}, our fine-tuned ResNet model exhibits superior performance on app identification and activity recognition with an accuracy of more than $90\%$, compared to the LSTM and transformer models with an accuracy of less than $90\%$. The transformer model performs better than LSTM due to the self-attention mechanism that can capture emanations over frequency and time. The superior performance of ResNet is mainly due to its revolutionary architecture, as it leverages residual learning
to address the vanishing gradient problem of DNN. As such, this design is particularly effective for extracting and classifying emanation features amidst diverse indoor environments.

\subsubsection{Impact of the number of frequency bands} Since the emanations spread across a wide frequency band, it is important to extract all the emanation spikes for accurate app identification and activity recognition. So, we need to explore the impact of the number of bands on app identification and activity recognition. 

\noindent \textbf{Method.} To do so, we mainly explore the frequency band between 585MHz and 626MHz with a bandwidth of 10MHz. As such, we have five frequency bands as the input of the fine-tuned ResNet for app identification and activity recognition. We report the accuracy of app identification and activity recognition when we use different numbers of frequency bands.

\noindent \textbf{Result.} Fig.~\ref{fig:app:accuracy:bands} shows the accuracy of the VR app identification and activity recognition over different numbers of frequency bands. As we can see, when there is only one frequency band, the accuracy of app identification is around 0.8. However, as the number of frequency bands increases, the accuracy of app identification increases to around 0.98. Since the app identification accuracy approaches one, we cannot further increase the accuracy when we use more than two frequency bands. This is because different apps can be accurately characterized by the emanation spikes that are periodically spread across a wide frequency band. This also indicates why we focus on frequency bands below 1 GHz. However, an app activity recognition accuracy is around 0.99, even with the emanations from one frequency band. This is because we leverage the over-time frequency characterization of the emanations to differentiate app activities. 

\begin{figure}
\centering
\captionsetup{width=0.23\textwidth}
\begin{minipage}{0.25\textwidth}
\centering
    \includegraphics[width=\linewidth]{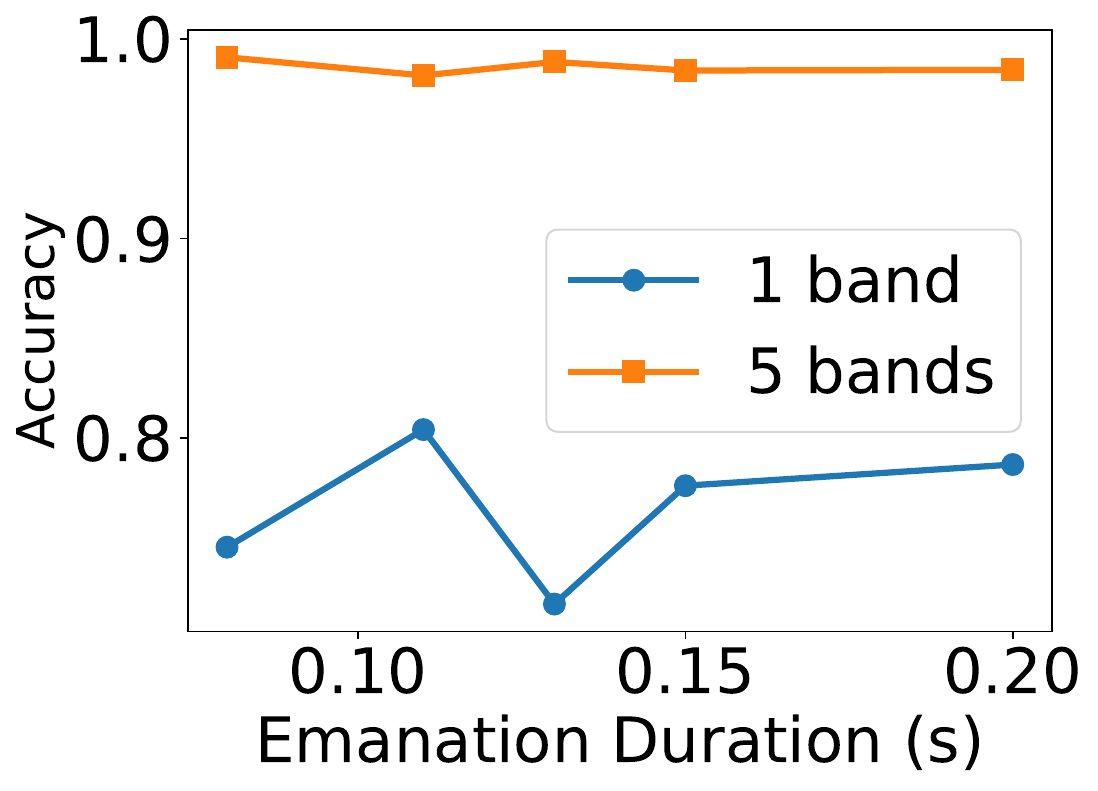}
    \caption{Accuracy of app identification across different emanation durations.}
    \label{fig:app:accuracy:duration}
\end{minipage}%
\begin{minipage}{0.25\textwidth}
  \centering
   \includegraphics[width=\linewidth]{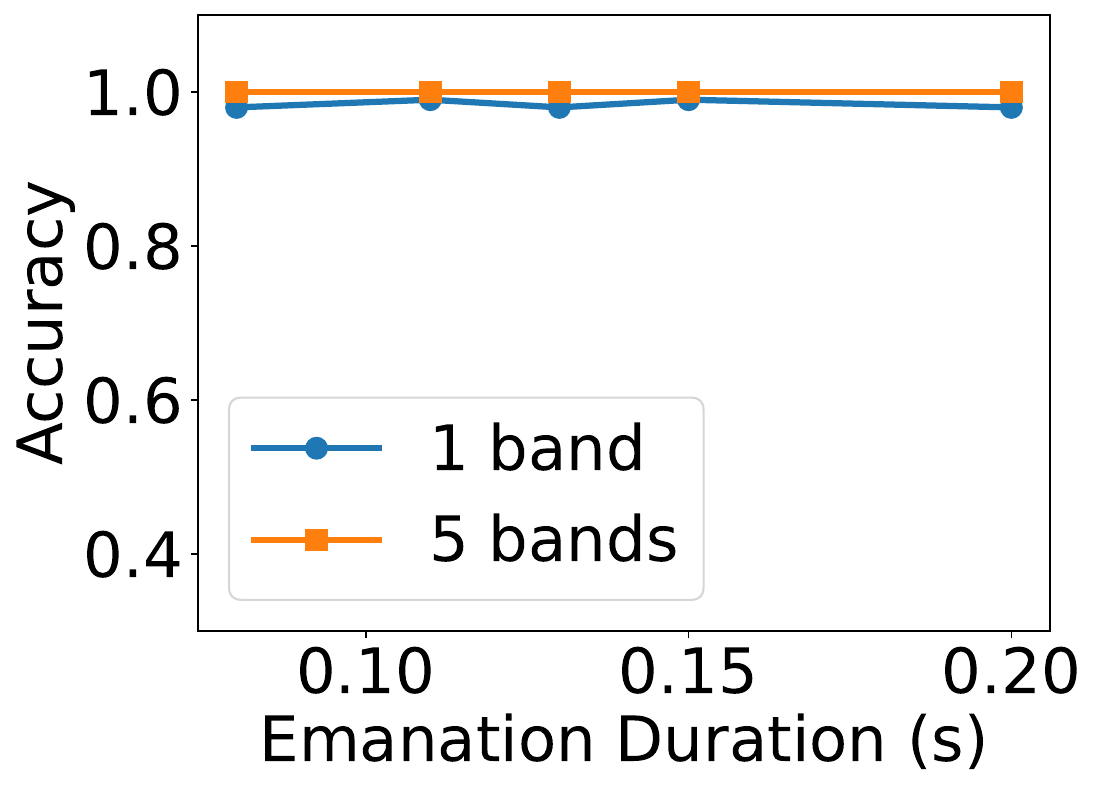}
    \caption{Accuracy of app activity recognition over emanation durations.}
    \label{fig:activity:accuracy:duration}
\end{minipage}
\end{figure}

\subsubsection{Impact of emanation signal duration} Emanation signals are sniffed with the software-defined radios over a period of time duration that can affect the frequency-domain emanation signal characterization. 

\noindent \textbf{Method.} To measure the impact of emanation signal duration on app identification and activity recognition, we derive the frequency-domain emanation representation over different emanation signal durations. The emanation signals are eavesdropped with USRP N210 at a sampling rate of 25 MHz for fine-grained spectrum characterization. 

\noindent \textbf{Result.} Fig.~\ref{fig:app:accuracy:duration} shows the accuracy of identifying the VR apps across different emanation durations. As we can see, the accuracy is around 0.98 across different emanation durations when we rely on emanations across five frequency bands. However, the accuracy is reduced to 0.76 across different emanation durations when we only rely on emanations over one frequency band. The accuracy does not vary significantly over the different durations. This is because the emanations are mainly characterized across frequency bands. The time-domain characteristics affected by the noise can be mitigated by our subtraction approach and the capability of the ResNet model. Moreover, the emanations are amplitude-modulated clock signals, whereby the emanation spikes can spread across a wide frequency band. As such, app identification across more frequency bands could provide better performance.

Fig.~\ref{fig:activity:accuracy:duration} shows the accuracy of app activity recognition across different emanation durations. As we can see, the accuracy of app activity recognition is around 0.99 when we rely on emanations spreading across one or five frequency bands for app activity recognition. This is because we use over-time frequency characterization of the emanations for app activity recognition, which aligns with what we have discussed about the impact of the number of frequency bands. Moreover, the accuracy does not change significantly over different durations due to the high sampling rate of 25 MHz, which can capture the detailed emanation characteristics.

\subsubsection{Impact of distance} The emanations can be attenuated over the air, which can affect the emanation reception strength at the eavesdropper. Intuitively, the longer the traversing distance, the weaker the reception of the emanation at the eavesdropper. 

\noindent \textbf{Method.} To evaluate the impact of the distance between the eavesdropper and the VR user's headset, we vary this distance to see the changes in the emanation spikes. We fixed the directional antenna's orientation at 90 degrees to the VR headset as shown in Fig.~\ref{fig:orientation:setup}.  As we can see in Fig.~\ref{fig:em:distance}, the USNR of the emanation spikes decreases over increasing distances. However, in this experimental evaluation, we plan to focus on the USNR variations of emanation spikes and the accuracy of app identification and activity recognition over longer distances. 

\begin{figure}
\centering
\captionsetup{width=0.23\textwidth}
\begin{minipage}{0.25\textwidth}
\centering
 \includegraphics[width=\linewidth]{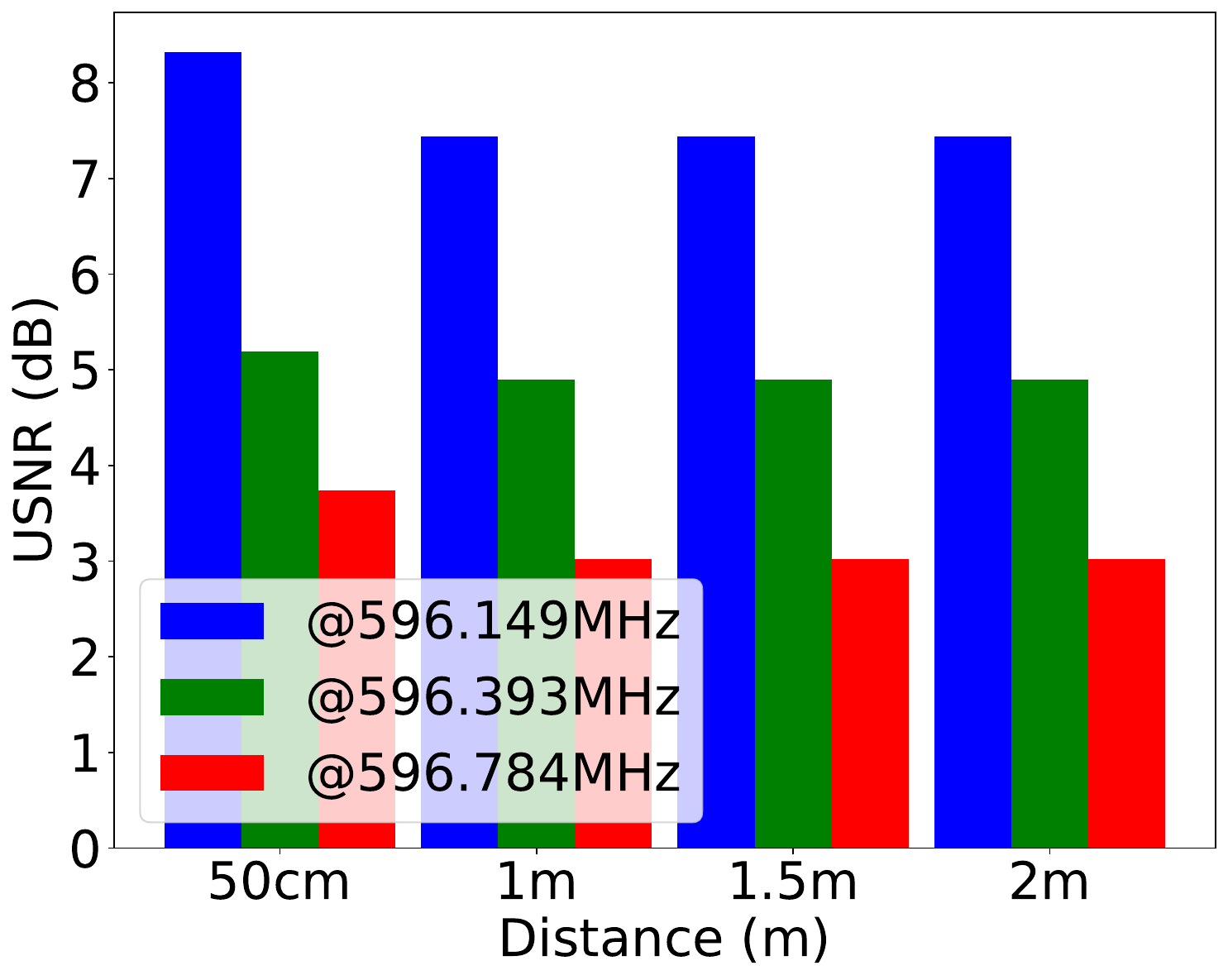}
    \caption{USNR of the emanation spikes over different distances.}
    \label{fig:distance:usnr}
\end{minipage}%
\begin{minipage}{0.25\textwidth}
  \centering
  \includegraphics[width=\linewidth]{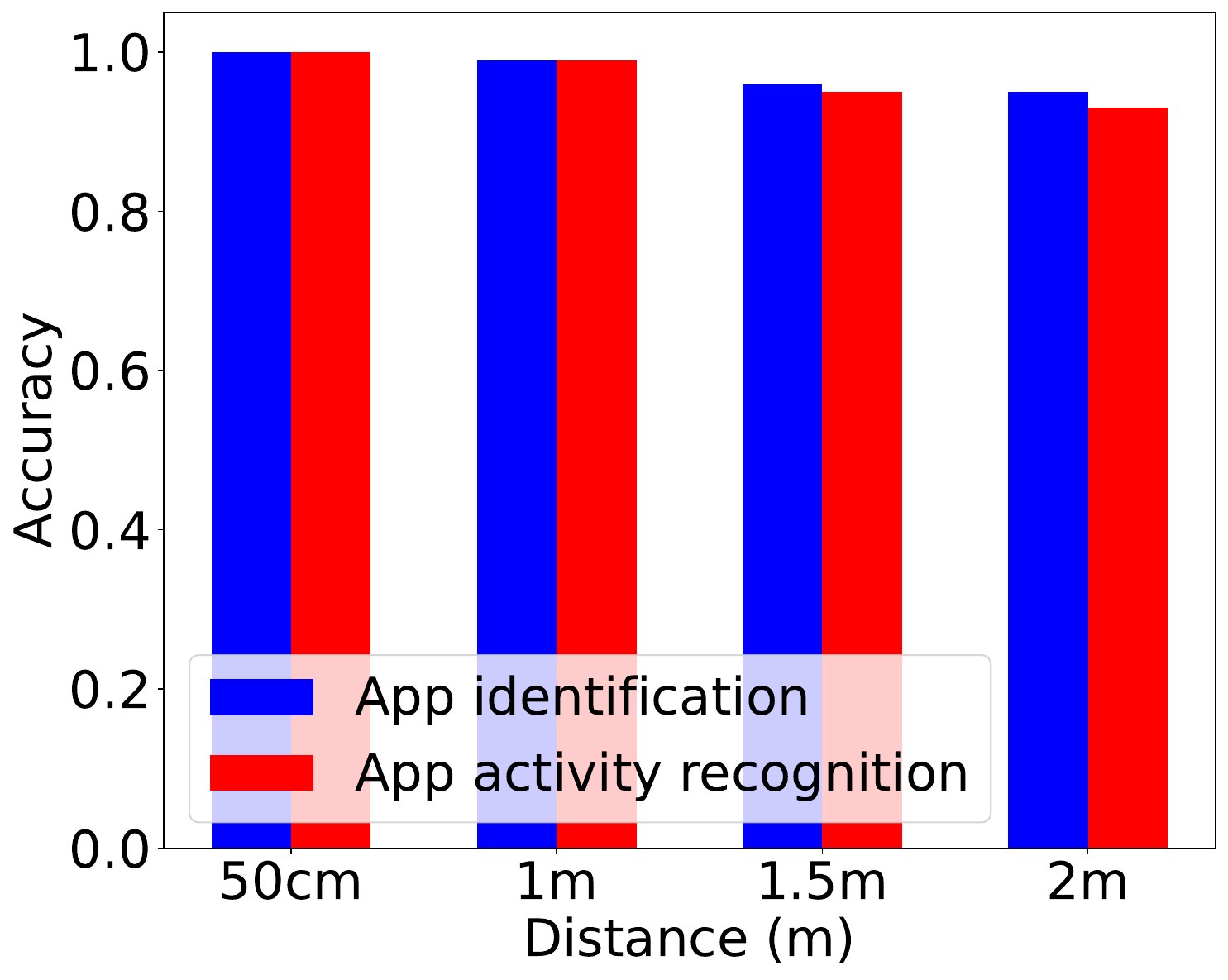}
    \caption{Accuracy of app identification and activity recognition over distances.}
    \label{fig:distance:accuracy}
\end{minipage}
\end{figure}

\noindent \textbf{Result.} Fig.~\ref{fig:distance:usnr} shows the USNR of the emanation spikes over the distances at frequencies of 596.149 MHz, 596.393 MHz, and 596.784 MHz. As we can see, the USNR decreases as the distance increases due to the over-the-air emanation attenuation. When the distance is over 1m, the USNR of the emanation spikes at the frequency of 596.149 MHz is around 7.5 dB. This is because the multipath effect becomes dominant over a longer distance, which can strengthen the received emanation signals due to the constructive signal addition.  Moreover, our subtraction approach can eliminate ambient wireless interference to enhance emanation detection. We also notice that the strength of the emanation spike decreases as the frequency increases, which complies with the characteristics of the spreading emanation signals over a wide frequency band, as we have discussed in the background section. To further improve the emanation strength, we can add a power amplifier at the eavesdropper. We also notice that different emanation spikes exhibit different strengths due to the spreading spectrum property of the emanations. Fig.~\ref{fig:distance:accuracy} shows the accuracy of app identification and activity recognition over different distances. As we can see, the accuracy of app identification and activity recognition slightly decreases as the distance increases. This is because the longer distance results in weaker emanation reception, which can be easily distorted by the ambient interference.

\subsubsection{Impact of eavesdropper-headset orientation} Since the eavesdropper uses the directional antenna to maximize the emanation signal reception, the orientation of the directional antenna to the VR headset could affect the emanation reception strength. Therefore, we evaluate the impact of orientation on the USNR of the emanation spikes.

\noindent \textbf{Method.} The eavesdropper's directional antenna can face the VR headset at different orientations. We mainly consider eight different orientations from $45^{\circ}$ to $360^{\circ}$ with $45^{\circ}$ separation. When the directional antenna faces the front end of the VR headset, it is 90 degrees. When the directional antenna faces the back end of the VR headset, it is 270 degrees. In this case, the emanations should penetrate the victim's head or be reflected off the objects in the environment for proper emanation sniffing. Note that the distance between the headset and the directional antenna is 1 meter across all the orientations. Then, we sniff the emanations to derive the USNR of the emanation spikes. As we can see, when the directional antenna faces the VR headset's front end, it is 90 degrees. We should receive the emanation signals with maximum strength. We focus on the emanation spikes within the frequency band between 590 MHz and 595 MHz.

\begin{figure}
\centering
\captionsetup{width=0.23\textwidth}
\begin{minipage}{0.25\textwidth}
  \centering
  \includegraphics[width=\linewidth]{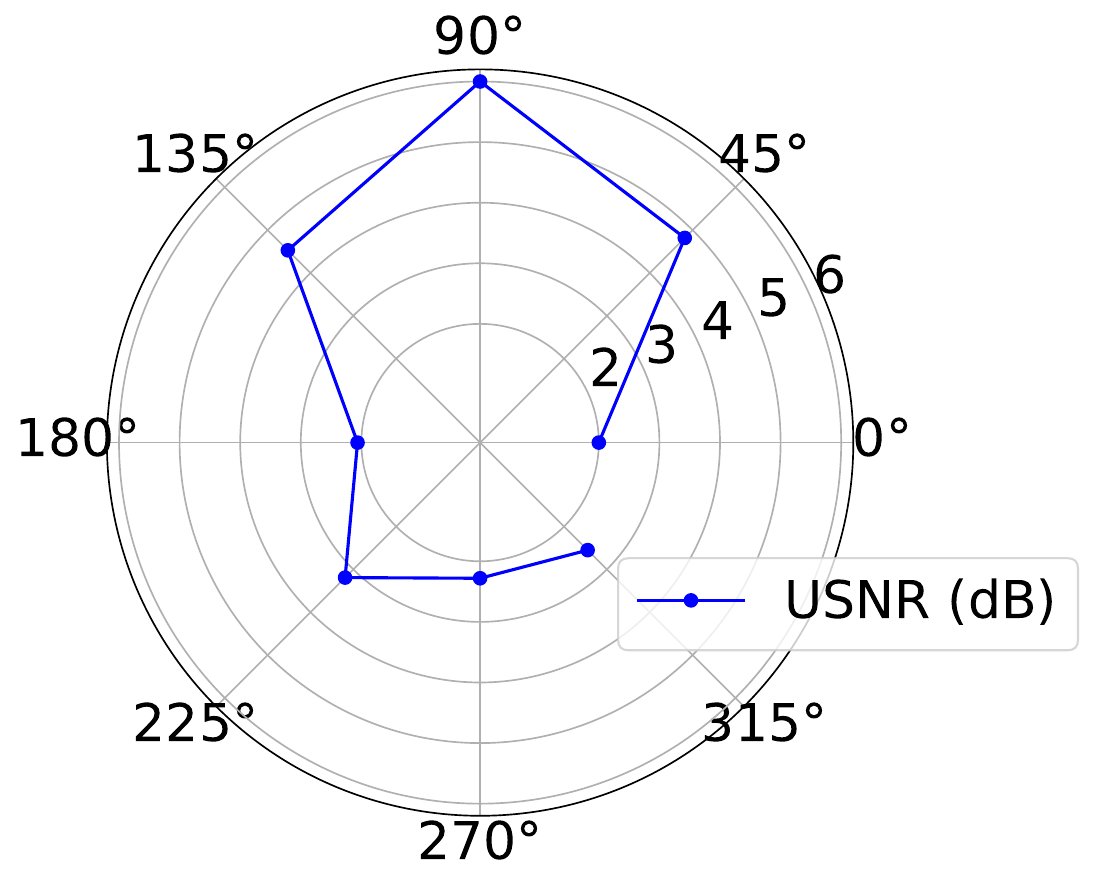}
    \caption{USNR of the emanation spikes when the eavesdropper's directional antenna faces the VR headset at different orientations.}
    \label{fig:orientation}
\end{minipage}%
\begin{minipage}{0.25\textwidth}
  \centering
  \includegraphics[width=\linewidth]{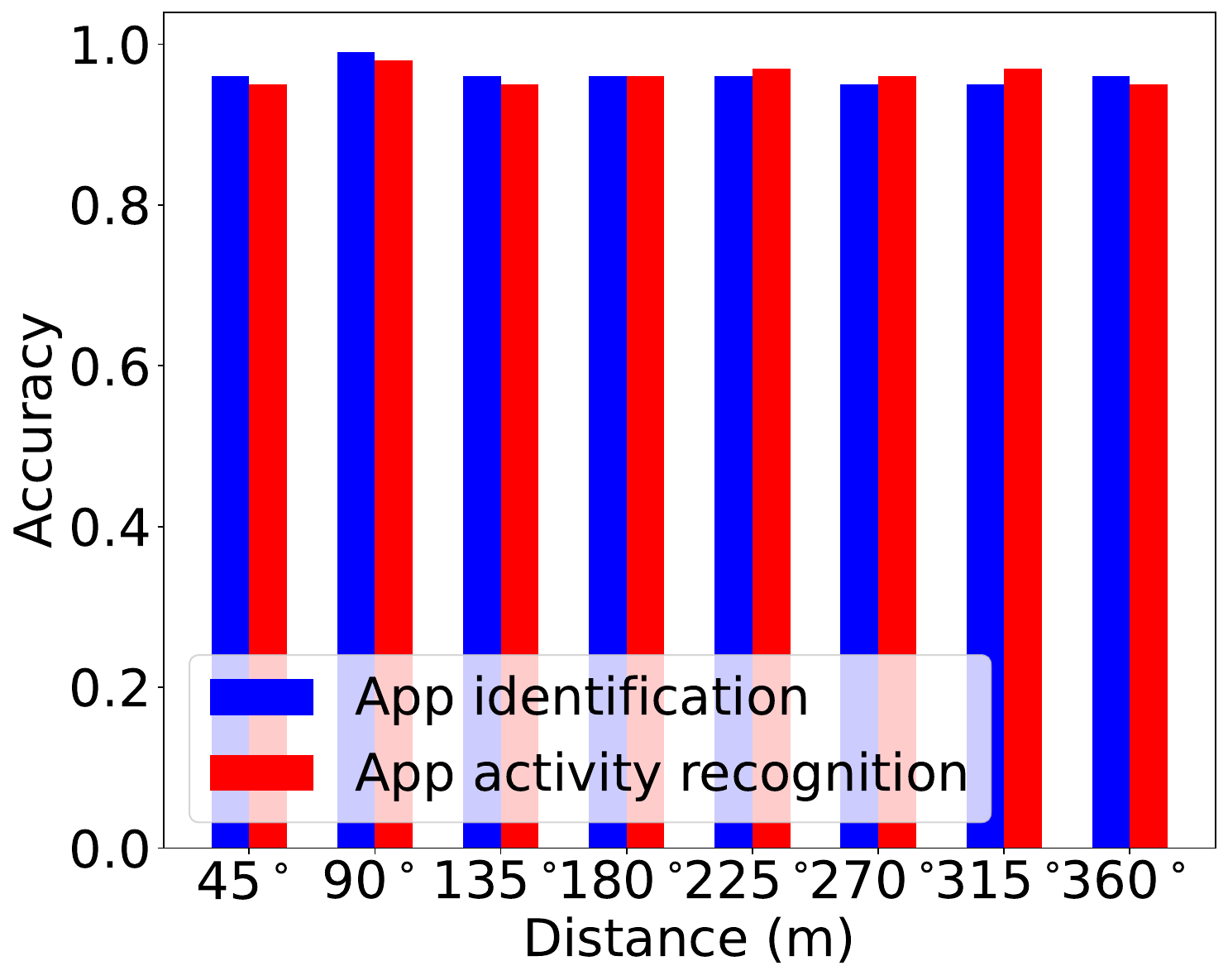}
    \caption{Accuracy of the app identification and activity recognition over different eavesdropper-headset orientations.}
    \label{fig:orientation:performance}
\end{minipage}
\end{figure}

\noindent \textbf{Result.} Fig.~\ref{fig:orientation} shows the USNR of the emanation spikes when the eavesdropper's directional antenna faces the VR headset at different orientations. As we can see, when the directional antenna faces the left and right sides of the VR headset, the emanation strength becomes the least (i.e., less than 1 dB) due to the poor directional antenna's orientation. When the directional antenna's orientation is between $45^{\circ}$ and $135^{\circ}$, the USNR becomes larger. At the orientation of $90^{\circ}$, the USNR is maximized. However, when the directional antenna's orientation is between $180^{\circ}$ and $360^{\circ}$, the emanation strength becomes weaker due to the head blockage. We find that the emanation strength becomes larger when the directional antenna is facing the front end of the VR headset. This is because the VR headset is worn on the victim's head. As such, when the directional antenna is facing the back end of the VR headset, the emanations are attenuated by the victim's head. Therefore, the attacker can use the directional antenna facing the VR headset's front end to maximize the emanation reception performance. Fig.~\ref{fig:orientation:performance} shows the accuracy of the app identification and activity recognition when the eavesdropper's directional antenna faces different orientations to the VR headset. As we can see, the accuracy of app identification and activity recognition is around 0.96 across the orientations. This is because \sysname leverages the machine learning models to characterize the signal pattern of the emanations that are not affected by the orientations. Moreover, even though the orientations could affect the USNR of the emanations, \sysname's system performance is not affected as long as the emanation spikes are characterized.

\section{Discussion}
\label{sec:dis}
\subsection{Countermeasures}


\noindent \textbf{EM prevention and jamming.} To defend against emanation-based privacy attacks on the VR headset, the straightforward idea is to prevent the emanation leakage from the VR headset through shielding or jamming the eavesdroppers. However, this is fundamentally difficult, as the emanations are automatically and unintentionally emitted from the VR headset, consisting of different kinds of IoT devices such as cameras and microphones. Shielding cannot fully prevent the emanation leakage. For example, we cannot fully shield the camera lens. Moreover, the jamming across the spectral spreading emanations could interfere with the in-progress legitimate wireless communication without knowing the eavesdropper's location.

\begin{figure}
\centering
\captionsetup{width=0.23\textwidth}
\begin{minipage}{0.25\textwidth}
\centering
    \includegraphics[width=\linewidth]{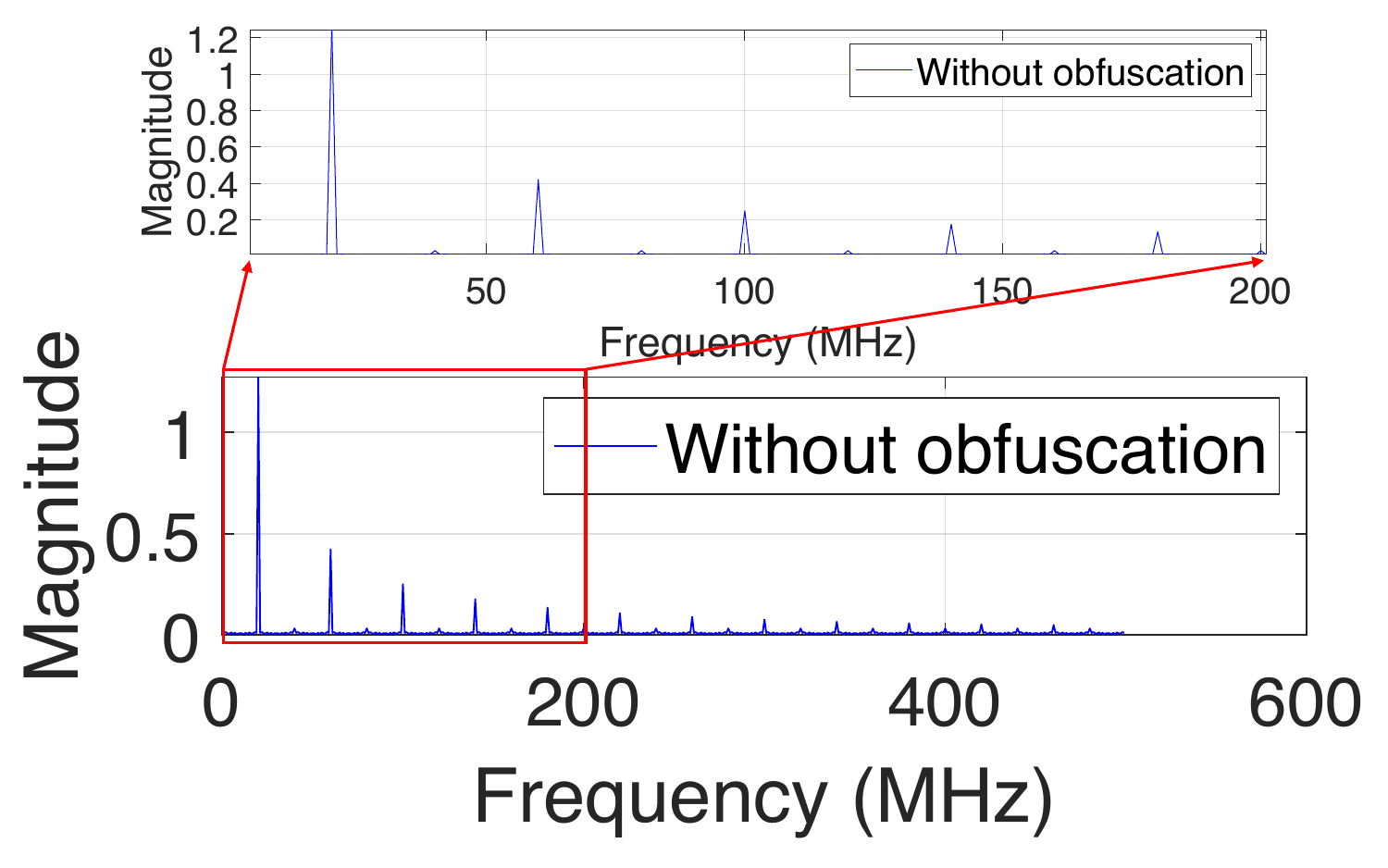}
    \caption{Frequency-domain emanations without EM obfuscation.}
    \label{fig:sim:no:obf}
\end{minipage}%
\begin{minipage}{0.25\textwidth}
  \centering
   \includegraphics[width=\linewidth]{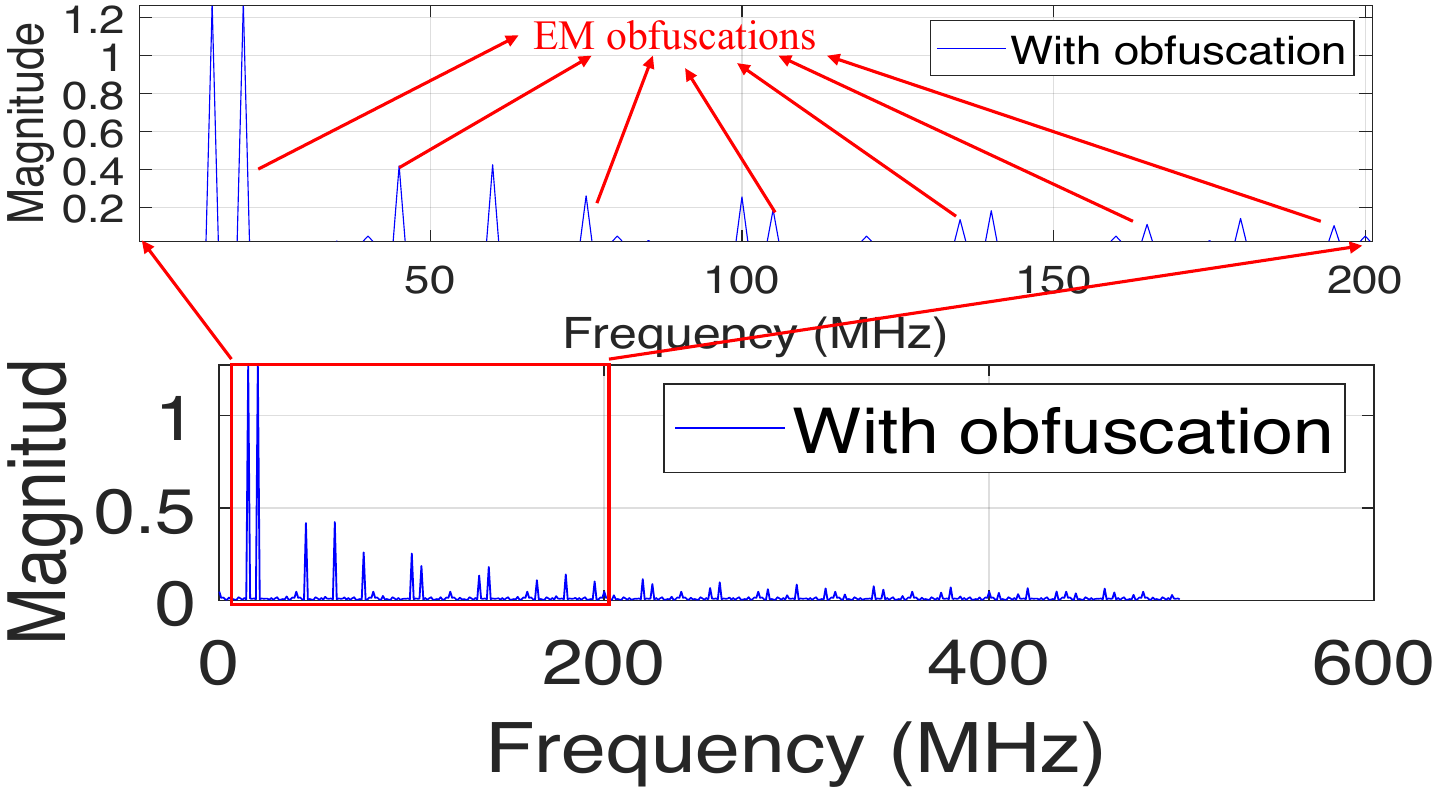}
    \caption{Frequency-domain emanations with EM obfuscation.}
    \label{fig:sim:obf}
\end{minipage}
\end{figure}
\noindent \textbf{EM obfuscation.} Another method to defend against this emanation-based attack on the VR headset is to obfuscate the emanations when they are emitted from the VR headset. As we have illustrated before, the emanations are amplitude-modulated clock signals that are affected by the computational activities on the VR headset. So, if we can obfuscate the computational activities on the CPU of the VR headset by running a daemon program, we can disable the emanation-based VR app identification and activity recognition. As shown in Fig.~\ref{fig:sim:no:obf}, we simulate the squared waves to represent the emanations in the frequency domain that are emitted from running VR apps without obfuscation. Fig.~\ref{fig:sim:obf} shows the frequency-domain emanations when the obfuscation is added through the simulated squared waves that can represent the emanations from running the daemon in the background. As we can see, the spectrum becomes different, and the simulated obfuscations are hard to suppress without prior knowledge of them, especially when the obfuscations are random. However, this can add extra overhead to the VR headset due to the daemon running in the background.

\noindent \textbf{Inference obfuscation.} Since our goal is to build the relationship between the emanations from the VR headset and the VR app identity and activity with machine learning models, we can obfuscate the machine learning model to disable this attack through adversarial learning, which highly relies on intelligent EM obfuscation in the above to generate adversarial examples in reality.

\subsection{Limitations and Future Work}

\noindent \textbf{Beyond app information inference.} In our study, we mainly focus on inferring the VR app characteristics. The emanations emitted from the VR headset can also reveal other important information. For example, we can use the emanations to infer the video contents and even reconstruct the video scenes in the VR headset, which requires us to build a relationship between the emanations and video contents with advanced machine learning models. This is because the video contents are related to the computational activities on the CPU of the VR headset, which can be revealed through emanations. Moreover, we focus on VR apps that are popular on the Meta Quest 3 and HTC VIVE XR Elite platforms. We believe that more types of apps and activities can still be accurately identified and recognized with well-trained machine learning models proposed in our work.

\noindent \textbf{Multi-user VR.} In our work, we mainly exploit the emanations to reveal the VR app characteristics from a single VR user for the targeted attack. Specifically, the attacker can use a directional antenna to target a specific VR user for app identification and activity recognition. When there are multiple VR headsets running different VR apps in a multi-user VR scenario, it is feasible to direct the directional antenna for emanation sniffing. However, it is difficult to differentiate the emanations from multiple VR headsets simultaneously. This is because the emanations from them spread across a wide frequency band and interleave with each other. The possible solution is to leverage the hardware imperfections of the VR headsets to differentiate the emanations from multiple headsets, which requires the VR headset to be characterized experimentally.

\noindent \textbf{Long-range eavesdropping.} In our work, we boost the emanation strength through the directional antenna and subtraction approach. To further boost the emanation strength, we can leverage electromagnetic interference (EMI) indicated in DeHiREC~\cite{zhou2023dehirec}. The basic idea of using EMI to boost the emanation strength is that the actively transmitted EMI could resonate with the hardware components in the VR headset to generate stronger emanations. To do so, we need to experimentally find out the EMI frequency that can efficiently excite the VR headset and deploy an extra transmitter to inject the EMI. Moreover, in our current work, we focus on the emanations from the VR headset with a directional antenna. To further enhance the system performance, we can also eavesdrop on the emanations from the hand controller as a complement to our existing settings.

\section{Related Work}
\label{sec:work}

\textbf{Virtual reality security.} Recently, virtual reality platforms have been widely studied to explore their security issues. However, the existing side-channel attacks on VR devices mainly focus on inferring the keystrokes~\cite{zhang2023s,al2021vr,ling2019know,luo2022holologger,luo2024eavesdropping,luo2020oculock,meteriz2022keylogging,slocum2023going,wu2023privacy,zhang2023s, yang2024can,wang2024gazeploit}, breathing and heartbeat patterns~\cite{zhang2023facereader}, VR user's location~\cite{farrukh2023locin}, facial muscle vibration-based authentication~\cite{zhang2023facereader}, motion-based VR user identification~\cite{nair2023unique} and speech extraction~\cite{cayir2025speak}, and network traffic-based keylogging attack~\cite{su2024remote} when the VR user is wearing the head-mounted display. These side channels include the acoustic emanations, the VR user's behavioral information (e.g., head movements), wireless channel state information, etc. For example, VR-Spy~\cite{al2021vr} leverages channel state information (CSI) from the wireless signals to infer keystrokes in the VR headset. Recently, Heimdall~\cite{luo2024eavesdropping} exploits the acoustic emanations from the VR controller to infer the keystrokes. INTRUDE~\cite{nguyen2024penetration} proposes to use the head movement information to infer the video types in VR headsets. A remote keylogging attack~\cite{su2024remote} is proposed to infer user-typed secrets using the network traffic side channel in multi-user VR applications. Papers in ~\cite{slocum2023going} and~\cite{nguyen2024penetration} explore the head motions to predict the user's keylogging information and VR videos, respectively.

Unlike these works, our work mainly leverages the automatic and unintentional electromagnetic emanations from the VR headset to infer the VR app identities and activities, which have never been explored in prior works.  Moreover, it is not clear that the previously explored side-channel information (e.g., head motions) could be related to the VR app identities and app activities.  However, the emanations can reveal the computational activities of the VR headset, which can be used for app identification and activity recognition. The prior works mainly focus on accurately extracting the frequency of the emanation spikes from the individual emanation source (e.g., camera, microphone) without considering the power spectral density and the over-time frequency properties of the emanations. Our work leverages the machine learning model to characterize the emanations from multiple emanation sources in the VR headset over time and frequency properties accurately.

\noindent \textbf{Electromagnetic side channel.} The electromagnetic side channel (i.e., emanation) has been studied for privacy attacks recently. Basically, the adversaries mainly eavesdrop on the side-channel information of the emanations emitted from the electronic device to infer private information such as keystrokes~\cite{jin2021periscope,vuagnoux2009compromising,wang2011analysis}, secrete keys~\cite{genkin2022lend,genkin2015stealing, guri2016usbee,de2023spectrem,goller2015side}, video, image, or audio contents~\cite{longeye,hayashi2014threat,tosaka2006feasibility,hayashi2016remote,lee2018information, chen2024eavesdropping, zhou2023dehirec, ni2023recovering,zhan2022graphics,liu2020screen,choi2020tempest}. However, these prior works have not exploited the characterized emanations for VR app identification and activity recognition. Moreover, these emanation-based privacy attacks do not fundamentally eliminate the interference from the ambient wireless communication signals. For example, EM Eye~\cite{long2024eye} infers the video content from the monitors or cameras. EM Eye mainly characterizes the emanations in the less-crowded spectrum using the amplifier to boost the emanation strength. Moreover, the emanations emitted from the large monitors are usually strong.  The emanations can also be used to detect the hidden or concealed IoT devices in the indoor environment for privacy protection~\cite{chaman2018ghostbuster,shen2021earfisher,stagner2011practical, li2024gpsbuster,liu2023camradar}. For example, RFScan~\cite{sun2025revealing} characterizes the emanations emitted from the hidden IoT devices for detection, fingerprinting, and localization, which mainly relies on signal processing to characterize the emanations.  IoTProsector~\cite{sun2023feasibility} designs a side-channel information-based inference approach and interactive tool for IoT enthusiasts to debug IoT devices based on the emanations and other side-channel information. However, IoTProsector mainly attaches an EM probe to the IoT devices for emanation detection and characterization.

Unlike previous works, our work moves one step further to reveal the characteristics of VR apps based on the emanations. Moreover, we cannot simply employ the prior techniques for VR identification and activity recognition due to domain-related concerns, which require over-time frequency-domain emanation characterizations and efficient machine learning models. 

\section{Conclusion}
\label{sec:cons}
In this paper, we thoroughly exploit the emanations emitted from the VR headset to infer the VR app identities and activities. To do so, we design \sysname, a system that can enhance emanation strength, suppress ambient wireless interference, and further characterize emanations with machine learning models for VR app identification and activity recognition. Our experimental evaluation reveals the efficiency of \sysname. We believe \sysname marks the first step in emanation-based information inference in VR. 


\bibliographystyle{ACM-Reference-Format}
\bibliography{reference}
\end{document}